\documentclass[aps,prx,twocolumn,superscriptaddress,noeprint,longbibliography, nofootinbib]{revtex4-1}

\usepackage{graphicx}
\usepackage{bm}
\usepackage{physics}
\usepackage{xcolor}
\usepackage{enumitem}
\usepackage{amsmath, amssymb}
\usepackage[normalem]{ulem}
\usepackage{natbib}
\usepackage{comment}

\newcommand{\be}{\begin{equation} }
\newcommand{\ee}{\end{equation} }
\newcommand{\ba}{\begin{eqnarray} }
\newcommand{\ea}{\end{eqnarray} }

\newcommand{\bit}{\begin{itemize}}
\newcommand{\eit}{\end{itemize}}

\graphicspath{{Figures/}}

\usepackage[colorlinks=true]{hyperref}
\hypersetup{citecolor = blue}

\newcommand{\dt}{\tau_{\rm f}}
\newcommand{\dth}{\tau_{\rm h}}
\newcommand{\heff}{H_{\rm eff}}

\begin{document}

\title{Prethermal stability of eigenstates under high frequency Floquet driving}

\author{Nicholas O'Dea}
\affiliation{Department of Physics, Stanford University, Stanford, CA 94305, USA}
\author{Fiona Burnell}
\affiliation{Department of Physics, University of Minnesota Twin Cities, MN 55455, USA}
\author{Anushya Chandran}
\affiliation{Department of Physics, Boston University, MA 02215, USA}
\author{Vedika Khemani}
\affiliation{Department of Physics, Stanford University, Stanford, CA 94305, USA}

\begin{abstract}
 Systems subject to high-frequency driving exhibit Floquet prethermalization, that is, they heat exponentially slowly on a time scale that is large in the drive frequency, $\dth \sim \exp(\omega)$. 
Nonetheless, local observables can decay much faster via energy conserving processes, which are expected to cause a 
rapid decay in the fidelity of an initial state.   
Here we show instead that the fidelities of \emph{eigenstates} of the time-averaged Hamiltonian, $H_0$, 
display an exponentially long lifetime over a wide range of frequencies -- even as generic initial states decay rapidly. 
When $H_0$ has quantum scars, or highly excited-eigenstates of low entanglement, this leads to long-lived non-thermal behavior of local observables in certain initial states. 
We present a two-channel theory describing the fidelity decay time $\dt$: the \emph{interzone} channel causes fidelity decay through energy absorption i.e. coupling across Floquet zones, and ties $\dt$ to the slow heating time scale, while the \emph{intrazone} channel causes hybridization between states in the same Floquet zone. Our work informs the robustness of experimental approaches for using Floquet engineering to generate interesting many-body Hamiltonians, with and without scars. 
\end{abstract}

\maketitle

\noindent \textbf{\textit{Introduction--}} 
Periodic driving is a powerful tool for engineering systems with interesting phases, including Hamiltonians with topological band structures or artificial gauge fields~~\cite{ 
Bukov:2015, Eckardt:2017, Oka:2019, Harper-Sondhi2020_review, Kitagawa-Demler_2010, lindner2011floquet, Goldman:2014, Khemani2016}. In the case of \emph{many-body} Hamiltonians, Floquet engineering relies on the existence of either many-body localization, or a `prethermal regime' in which the heating due to the drive is exponentially suppressed at high driving frequencies~\cite{ Abanin-Huveneers2015_exponentially,Mori-Saito2016_rigorous, Kuwahara-Saito2016_floquet,Abanin-Huveneers2017_rigorous}. In the latter case, this implies that gapped ground states with quantum order can be engineered by driving, and can survive for exponentially long times.

More recently, Floquet engineering has also been used to experimentally realize and stabilize Hamiltonians with \emph{quantum scars}, which are atypical highly-excited eigenstates with low entanglement and outlier expectation values~\cite{shiraishi_systematic_2017, bernien_probing_2017, Serbyn-Papic2021_review,Moudgalya-Regnault2021_review,Chandran-Moessner2022_review, bluvstein_controlling_2021, maskara_discrete_2021,  Mukherjee_collapse:2020, Mukherjee_dynamics:2020}. These states exist throughout the many-body spectrum and can induce striking dynamics, including quantum coherent oscillations in `hot' initial states at infinite temperature.
To explain the efficacy of Floquet engineering in this context where \emph{highly-excited} eigenstates are of interest, bounds on heating are not sufficient because the eigenspectrum is exponentially dense (in system size) near the states of interest. Instead, one additionally requires bounds on \emph{eigenstate fidelity} which control the lifetimes of individual eigenstates, for which there are few existing results.  Developing such an understanding is important as the origin, properties and robustness of quantum scars are subjects of active study.

\begin{figure}[h]
\includegraphics[width=\linewidth]{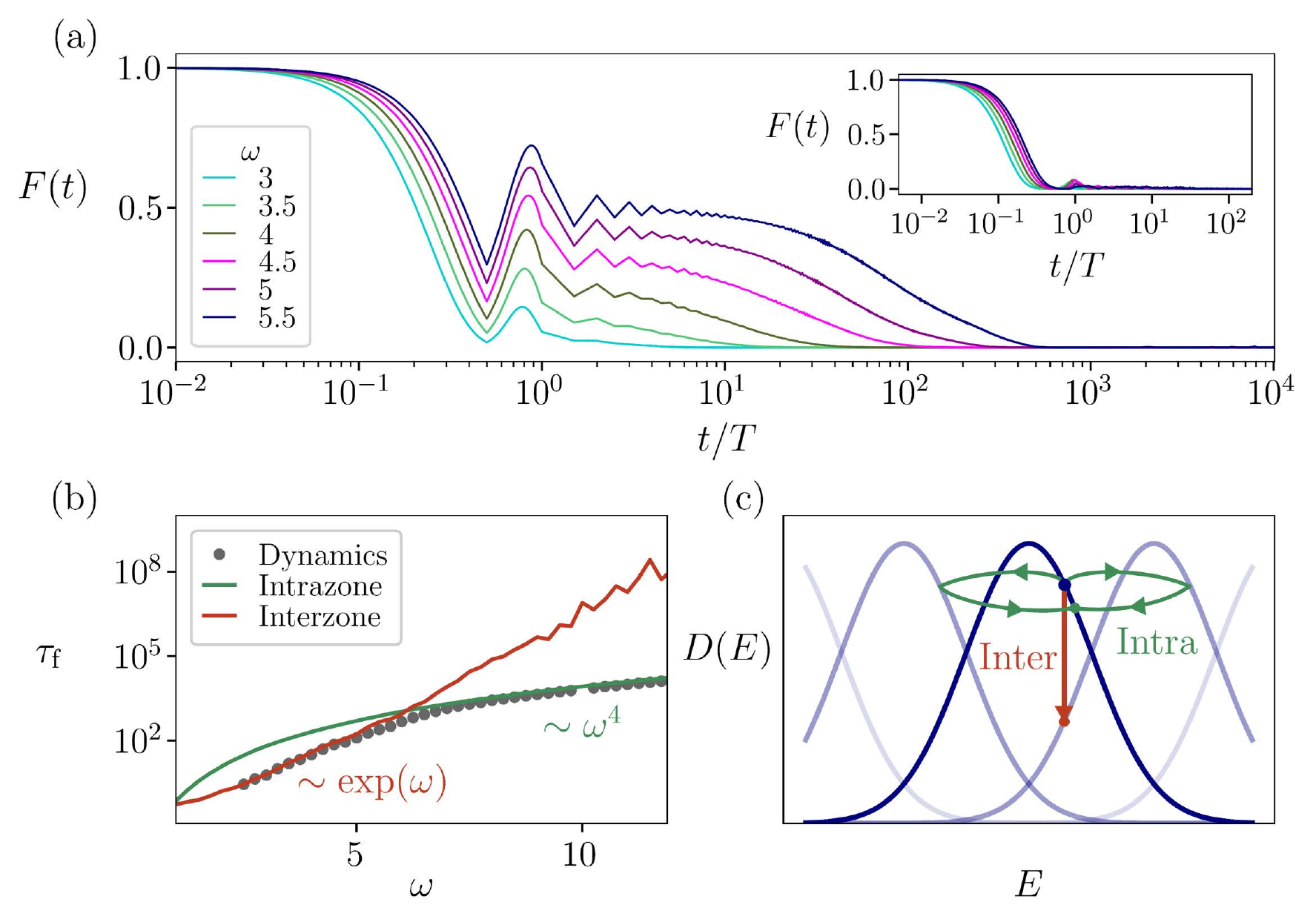}
 \caption{(a) The fidelity against time, $F(t)$, for a highly-excited eigenstate of $H_0$ at zero energy density driven by $H(t)$ (Eq.~\eqref{eq:periodic_H}). The fidelity is long-lived, and the lifetime shows strong frequency dependence. Inset shows the rapid decay of $F(t)$ for a zero energy density product state $\ket{00...0}$ (not an eigenstate of $H_0$). (b) Comparison of the numerically extracted fidelity decay timescale $\dt$ with the interzone and intrazone FGR predictions~\cite{Supplementary}, demonstrating the crossover between the two regimes. The data is for the same eigenstate as in (a). (c) Schematic showing the density of states of $H_0$ vs energy $E$ in a replicated zone scheme. The arrows show the dominant interzone and intrazone processes causing fidelity decay.  Parameters: model in Eq.~\eqref{Eq:ModelH}, $L=14$, $F(t)$ evaluated at stroboscopic times after one period.}
\label{fig:main1}
\end{figure}

To address this gap in the literature, we identify mechanisms for long fidelity lifetimes of arbitrary eigenstates in Floquet engineering approaches.
The practical target for these approaches is the time-averaged Hamiltonian $H_0$.  Here, we numerically and analytically quantify
the time-dependent
fidelity~\cite{gorin_loschmidt_2006, zarate-herrada_generalized_2023} of a generic (possibly highly-excited) eigenstate $|\psi\rangle$ of $H_0$:
\begin{equation}
    F(t) = |\langle \psi| U(t) |\psi \rangle|^2. 
\end{equation} 
under the Floquet time evolution operator $U(t)$ generated by the full time-dependent Hamiltonian $H(t)$.

This fidelity will decay through two channels: (i) heating and (ii) dressing of the dynamics relative to that generated by $H_0$. The timescale for heating is $\dth \geq \exp(\omega/J)$, where $J$ is a typical instantaneous local energy scale in the problem, and $\omega$ is the drive frequency. Intuitively, the system requires $O(\omega/J)$ local rearrangements to absorb one quantum ($\omega$) of energy; beyond the heating time, the system can no longer be described by a quasi-local effective Hamiltonian, $\heff$. While $\heff$ describes the local dynamics in the prethermal regime, $H_0$ is only the leading term in this Hamiltonian, which can be constructed order-by-order in a high-frequency expansion in $(1/\omega)$.  The difference between $\heff$ and $H_0$ leads to a dressing of $H_0$ that is $O(1/\omega)$, and a naive Fermi Golden Rule (FGR) calculation thus predicts a decay timescale for the fidelity $\dt \sim \omega^2$,  parametrically shorter than the heating time.

Surprisingly, as shown in Fig.~\ref{fig:main1}(a), we find that eigenstates of $H_0$ survive for much longer over a wide range of frequencies. The fidelity curves are roughly linearly spaced on a semi-log scale, pointing to an \emph{exponential-in-frequency} time-scale for the fidelity decay, similar to $\dth$. These long lifetimes are only observed for eigenstates of $H_0$; in contrast, the inset shows that the fidelity of a generic state rapidly decays over just one driving period. In ~\cite{Supplementary}, we show heating and fidelity data for both generic initial states and eigenstates of $H_0$; while all states show exponentially slow heating, only eigenstates of $H_0$ show long-lived fidelity.  Interestingly, we find that scarred eigenstates survive for \textit{even longer} than non-scarred eigenstates of $H_0$ at similar energies. Initial states with large overlaps on these scars can be experimentally prepared, and we show that dynamics of local observables then inherit the stability of eigenstates. Our work opens the door for novel Floquet engineering approaches for generating many-body Hamiltonians with interesting excited eigenstates, including (but not limited to) quantum scars.

\noindent \textbf{\textit{Two channels for fidelity decay--}}
To explain these unexpectedly long fidelity lifetimes, we analyze two channels that contribute to fidelity decay (see Fig. \ref{fig:main1}b): an ``intrazone channel"  incorporating the effects of dressing, and an ``interzone channel" that causes heating.  Intrazone processes are well-captured by a perturbative high frequency expansion (solid green line), while heating processes, which are non-perturbative in $\omega$, are well-captured by second-order time-dependent perturbation theory (TDPT) (solid red line.).  \emph{Both} channels yield decay timescales which are parametrically longer than the naively expected $\dt \sim \omega^2$.  

A striking feature of Fig.~\ref{fig:main1}(b) is that for a wide range of frequencies, the interzone channel dominates and gives $\dt \sim \exp(\omega)$. At larger frequencies, the higher-order intrazone channel determines the observed $O(\omega^4)$ lifetime. 
In practice, however, the exponentially large interzone fidelity decay timescale might be all that is empirically accessible, since real experiments have to be careful to not excite high-energy degrees of freedom that lie outside the scope of the original isolated system, such as higher bands in lattice models.  Similarly, in numerics finite system sizes limit the energy bandwidth and hence the range of accessible frequencies.

We consider periodically driven systems in $d$ spatial dimensions of the form
\begin{equation}\label{eq:periodic_H}
    H(t) = H_0 + \lambda V g(t), 
\end{equation}
where $g(t) = g(t+T)$ is a periodic function with period $T=2\pi/\omega$ which time-averages to zero. The linear dimension is $L$, and the initial state $|\psi \rangle$ is an eigenstate of $H_0$.
We consider short-ranged models $H_0 = \sum_i h_i, V=\sum_i V_i$, where $h_i, V_i$ have support near site~$i$ and set the local energy scale to $J$. The stroboscopic dynamics is governed by the Floquet unitary $U_F = \mathcal{T} \exp\left(-i \int_0^T H(t) dt\right)$ which time-evolves the system by one period and defines the Floquet Hamiltonian $H_F$ via $U_F \equiv e^{-i H_F T}$. 
In the prethermal regime $t \lesssim \dth$,  the stroboscopic dynamics is well-described by a quasi-local effective Hamiltonian~\cite{Abanin-Huveneers2015_exponentially, Kuwahara-Saito2016_floquet} 
\begin{equation}\label{eq:highfreqexp}
    \heff = H_0 + \sum_{m=1}^n \frac{H^{(m)}}{\omega^m} 
\end{equation}
where $H^{(m)}$ is the $m^{th
}$  order term in the high-frequency Magnus expansion, and the series is truncated at an optimal order $n \propto \omega/J$.  Because $\heff$ is almost conserved, its eigenstates are theoretically guaranteed to show slow dynamics up to times $t \lesssim \dth$. 
However, $\heff$ does not include the effects of heating, as the heating rate $\dth^{-1} \sim e^{-\omega/J}$ is non-perturbative in powers of $(1/\omega)$.  The process of Floquet heating entails the system resonantly exchanging energy with the drive in quanta of size $\sim \omega$, which is visualized as a drive mediated coupling across ``Floquet zones" representing copies of the eigenspectrum of $H_0$ displaced by integer multiples of $\omega$ (Fig. \ref{fig:main1}(c)).

Thus, the quasi-local and approximately conserved $\heff$ describes the dynamics induced by the drive within a Floquet zone (intrazone), while the heating process is described by coupling between zones (interzone). The latter is reflected in the non-perturbative, non-local character of $H_F$: the difference between $U_F=e^{-iH_FT}$ and $e^{-iH_{\rm eff}T}$ is the heating process across Floquet zones visible on times $t> \dth$. 

To find compact, analytical forms for the fidelity decay rate in terms of matrix elements in the $H_0$ eigenbasis, we use time-dependent perturbation theory (TDPT) and the assumptions of Fermi's golden rule (FGR)~\cite{Supplementary}. The Dyson series for the time-evolution operator $U_I(t)$ in the interaction picture leads to a perturbative expansion of $\log(F(t))$ in powers of $\lambda$:
\begin{align}\label{eq:mainDyson}
  \log (   F(t) ) &= 2\Re [ \log(\langle \psi |U_I(t)| \psi \rangle)] \nonumber \\ &= \sum_{n=0}^\infty \lambda^n G^{(n)}(t)
\end{align}
Note that the series in Eq.~\eqref{eq:mainDyson} contains both perturbative and non-perturbative contributions in $\frac{1}{\omega}$.  The time evolved state $U(t)|\psi \rangle$ generically deviates from $|\psi \rangle$ on all sites; the fidelity is consequently exponentially small in $L^d$, so that its logarithm has an extensive series expansion. The log fidelity decay rate $\propto L^d$ is consistent with a system-size independent decay rate of local observables. We confirm this in~\cite{Supplementary} via numerical study of the quantity $||\rho(t)-\rho(0)||$, the change in the local reduced density matrix, which bounds changes in local observables and which we find to be asymptotically system-size independent. 

The leading contributions to the intrazone and interzone processes are respectively derived from the $O(\lambda^2)$ and $O(\lambda^4)$ terms of the $\log(F(t))$ TDPT series in Eq.~\eqref{eq:mainDyson}. Expressions for general drives are provided in~\cite{Supplementary} which, for a sinusoidal drive, $g(t) = \sin(\omega t)$ take the form:
\begin{equation}
    \label{eq:interzone_sin}
    \begin{split}
    R^{\text{inter}}_\psi &= \frac{\pi \lambda^2}{2}  \left(\mathcal{A}^{V}_\psi(\omega) + \mathcal{A}^{V}_\psi(-\omega) \right)\\
    \end{split}
\end{equation}
\begin{equation}
    \label{eq:intrazone_sin}
    R^{\text{intra}}_\psi = \frac{\pi \lambda^4}{8 \omega^4} \mathcal{A}_\psi^{[V,[H_0,V]]}(0)
\end{equation}
Here, $\mathcal{A}_\psi^{O}(\omega) = \sum_{n} |O_{\psi n}|^2 \delta( (E_n - E_\psi)-\omega)$ denotes the spectral function of an operator $O$ in an eigenstate $|\psi\rangle$ of $H_0$. When $O$ is an extensive sum of local operators, the spectral function is exponentially suppressed at large frequencies, $\mathcal{A}^{O}_i(\omega) \sim L^d e^{-|\omega|/J}$.~\footnote{In one dimension, $\mathcal{A}^{O}_\psi(\omega)$ is bounded above by a value of order $L e^{-|\omega|\log(|\omega|)/J}$~\cite{Abanin-Huveneers2015_exponentially} The extra suppression from the $\log(\omega)$ in the argument of the exponential does not significantly change any results.} 

The interzone rate $R^{\text{inter}}_\psi$ is exponentially suppressed in frequency via the spectral function $\mathcal{A}_\psi^{V}(\pm \omega)$. The leading process contributing to $R^{\text{inter}}_\psi$ couples the state $\psi$  to states in  neighboring Floquet zones (Fig.~\ref{fig:main1}(c)). This happens at first order in the driving, $O(\lambda)$, which squares to give an $O(\lambda^2)$ contribution to $F(t)$. 
Interzone coupling is also responsible for energy absorption in multiples of the frequency $\omega$; indeed, for $g(t) = \sin(\omega t)$, the heating rate is~\cite{Supplementary},
\begin{equation}
    \label{eq:heating_sin}
R^{\text{heat}}_\psi = \frac{\pi \lambda^2}{2} \omega  \left(\mathcal{A}_\psi^{V}(\omega) - \mathcal{A}_\psi^{V}(-\omega) \right),    
\end{equation}
which involves the same spectral functions as in Eq.~\eqref{eq:interzone_sin}.

In contrast, the leading contribution to the intrazone rate $R^{\text{intra}}_\psi$ is a second-order process which couples $\psi$ to a nearby-in-energy state $\psi_f$ in the same Floquet zone, mediated by a virtual hop to a state $\psi_k$ in a neighboring Floquet zone (Fig.~\ref{fig:main1}(c)). This process involves terms of the form $\sum_k  \lambda^2 \langle \psi_f|V| \psi_k \rangle \langle \psi_k |V| \psi\rangle/ (E_k - E_f + \omega)$ when $\psi_k$ belongs to the zone shifted by $+\omega$. Due to the decay of spectral functions with energy differences, the matrix elements are largest for processes where $E_\psi \approx E_k \approx E_f$ (Fig.~\ref{fig:main1}(c)), which gives a $1/\omega$ contribution in the denominator. Upon squaring, the intrazone rate is thus naively expected to scale as $\frac{1}{\omega^2}$.  However, in TDPT a cancellation between terms involving virtual transitions to neighboring Floquet zones, i.e. spectra shifted by $\pm \omega$ (Fig.~\ref{fig:main1}c), leads instead to a dependence scaling as $\frac{1}{\omega^4}$~\cite{Supplementary} and controlled by the spectral function of $[V, [H_0,V]$. Heuristically:
\begin{align}\label{Eq:heuristic_cancellation}
    &\sum_k \frac{\lambda^2 \langle \psi_f|V |\psi_k \rangle \langle \psi_k |V| \psi\rangle}{ (E_k - E_f + \omega)} + \frac{\lambda^2 \langle \psi_f|V |\psi_k \rangle \langle \psi_k |V| \psi\rangle}{ (E_k - E_f - \omega)} \nonumber \\
    &\approx \frac{2\lambda^2}{\omega^2} \sum_k \langle \psi_f|V |\psi_k \rangle \langle \psi_k |V| \psi\rangle (E_k - E_f)\nonumber \\
    & \approx  \frac{\lambda^2}{\omega^2} \langle\psi_f|[V, [H_0,V]]|\psi\rangle
\end{align}
 where we have kept the leading $1/\omega$ contribution and used $E_\psi \approx E_k \approx E_f$. This squares to give an $O(\lambda^4/\omega^4)$ leading contribution to $R^{\text{intra}}_\psi$.  The cancellation above relies on the fact that the matrix elements coupling $\psi$ to states in neighboring zones are guaranteed to be the same, since the zones are just shifted copies of the spectrum of $H_0$.

The intrazone contribution in Eq.~\eqref{eq:intrazone_sin}, which is perturbative in $\frac{1}{\omega}$, can also be obtained from TDPT applied to the high-frequency Floquet-Magnus expansion in Eq.~\eqref{eq:highfreqexp}. In this latter method, we truncate the high-frequency expansion of the Floquet-Magnus Hamiltonian to order $O(\frac{1}{\omega^2})$, and treat the dynamics as a quench to this time-independent Hamiltonian at $t=0$ via Eq.~\eqref{eq:mainDyson}~\cite{Supplementary}.  In the Floquet-Magnus treatment, the $1/\omega^4$ dependence arises because the spectral function of $H^{(1)}\propto [H_0,V]$ in Eq.~\eqref{eq:highfreqexp} has a soft gap, $\mathcal{A}_\psi^{H^{(1)}}(0)=0$, such that $H^{(1)}$ does not directly contribute to the FGR rate. The next order term $H^{(2)} \propto [V, [H_0,V]]$ is $O(\frac{1}{\omega^2})$, and contributes the observed $O(\frac{1}{\omega^4})$ rate via FGR. Note that drives with multiple non-commuting terms, for example $V_A \cos(\omega t)+V_B\cos(\omega t+\phi)$  can induce a term  $ H^{(1)} \propto \frac{[V_A, V_B]}{\omega}$ in the first order of the Floquet-Magnus expansion, which can lead to an asymptotically larger intrazone rate  $O(\frac{1}{\omega^2})$. Nevertheless, there can remain an intermediate frequency window in which the decay is dominated by interzone processes and $\tau_f$ increases exponentially with frequency. 

We expect the larger rate to set the timescale for the fidelity decay, $\dt \sim \frac{1}{R^{\text{inter}}+R^{\text{intra}}}$. Thus at large frequency, there is a crossover from interzone to intrazone dominated fidelity decay. The crossover frequency is $O(1)$ but can be made parametrically large by reducing $\lambda/J$~\cite{Supplementary}. In general, weak, high-frequency driving is most amenable to observing prethermal stability for eigenstates of $H_0$. 

\textbf{\textit{Numerics--}}
We test our two-channel picture through exact numerics on spin chains. Consider the following spin-$1$ scarred $H_0$ and driving protocol:
\begin{equation}\label{Eq:ModelH}
\begin{aligned}
    &H_0 = \sum_i h_i \\
    & h_i =  S^x_i S^x_{i+1} + S^y_i S^y_{i+1} + .1 \left( S^x_i S^x_{i+3} + S^y_i S^y_{i+3} + (S^z_i)^2\right ) \\
    &H(t) = H_0 +  \lambda \text{Sgn}\left(\sin(\omega t) \right) \sum_i S^x_i
\end{aligned}
\end{equation}
Numerics are performed with $\lambda = 0.5$. 

Without the drive, $H_0$ is the spin-1 XY model which conserves total $z$-magnetization and has a tower of quantum scarred eigenstates \cite{schecter_weak_2019,mark_unified_2020, odea_tunnels_2020}, $\left(\sum_{i=1}^L (-1)^i (S^+_i)^2\right)^n |---....---\rangle$ for $n=0,1,...,L$. There is one such scar state in each $S^z=-L, -L+2, ..., L$ sector, with quasi-momentum $0$ or $\pi$ depending on whether $n$ is even or odd. The scar states have energy density $.1$, close to the infinite temperature energy density of $.0\overline{6}$. The scar states are not eigenstates in the presence of the $S^x$ drive.

\begin{figure}[tb]
\includegraphics[width=\linewidth]{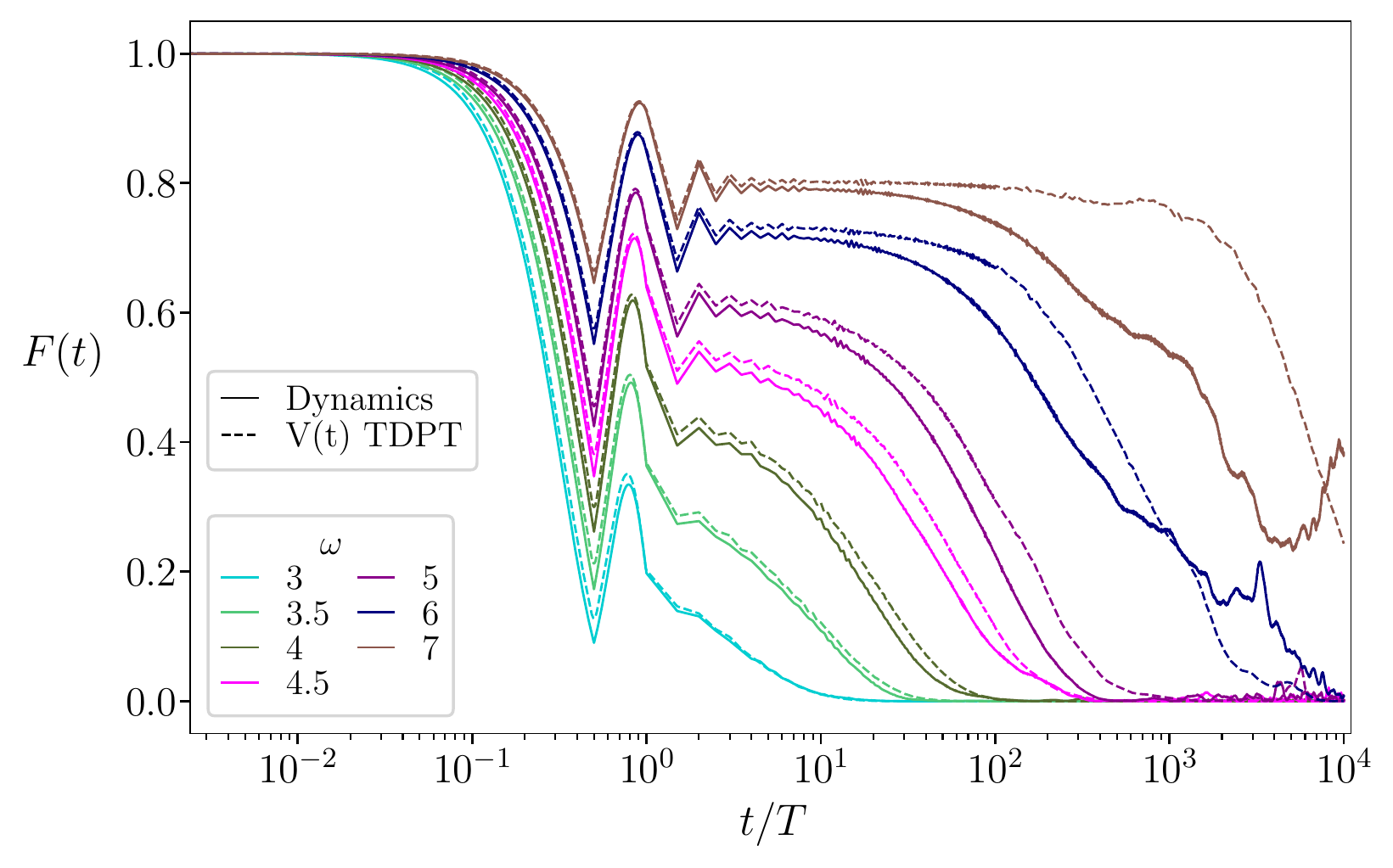}
\caption{Comparison of numerically exact fidelity dynamics under $H(t)$ (solid line) with 2nd-order TDPT (dotted line) for different drive frequencies $\omega$. TDPT predicts the entire curve $F(t)$ well for $3 \leq \omega \lesssim 5$.  Parameters: $L=14$, eigenstate of $H_0$ in the $S^z=0$, momentum $k=0$, inversion antisymmetric, spin-flip antisymmetric sector next to the scar state in energy. } 
\label{fig:main2}
\end{figure}

We implement a box drive with multiple harmonics. The resulting interzone and intrazone fidelity decay rates are obtained using the same methods as for Eqs.~\eqref{eq:interzone_sin} and \eqref{eq:intrazone_sin} and give qualitatively similar results~\cite{Supplementary}. 
Fig.~\ref{fig:main1}(a) (discussed previously) plots $F(t)$ for a zero energy density eigenstate. Fig.~\ref{fig:main1}(b) compares the numerically extracted decay timescale $\dt$ to the two-channel FGR predictions. We discuss the numerical extraction of the two-channel rates and $\tau_f$ in \cite{Supplementary}. The predicted rates match the data well, with the exponentially suppressed interzone rate dominating at small frequencies and crossing over to the polynomially suppressed intrazone rate at larger frequencies,  $\omega \gtrapprox 6$. 

Furthermore, TDPT provides more than just a FGR rate - it can also produce quantitatively accurate fidelity decay curves. Fig.~\ref{fig:main2} compares the numerically exact fidelity for an eigenstate of $H_0$ in the $S^z=0$ sector with energy close to that of the scar state to the results from second-order TDPT applied to $V(t)$ for different drive frequencies (see Eq.~\eqref{eq:mainDyson}). The two compare well at all times for small frequencies, and at early times for large frequencies. 

\begin{figure}[tb]
\includegraphics[width=\linewidth]{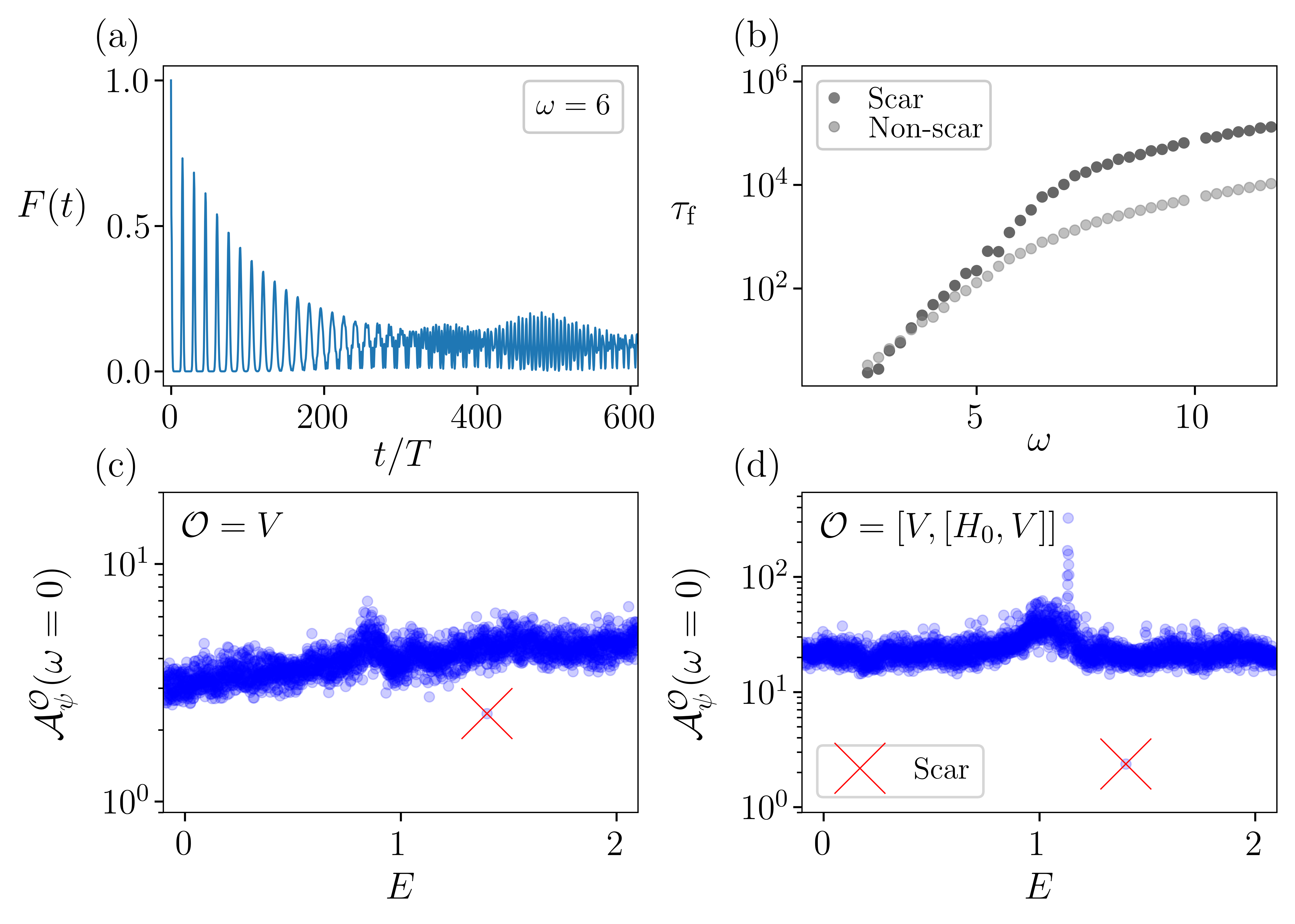}
\caption{(a) Fidelity dynamics of the scarred product state $\bigotimes_{i=1}^L \frac{|+\rangle_i + (-1)^i |-\rangle_i}{\sqrt{2}}$ for time evolution with $H(t) + h S^z$ with $L=14$ and $h=.2$. The fidelity exhibits long-lived oscillations. The remaining panels consider the scar state in the $S^z=0$ sector for comparison to other eigenstates in the same sector of $H_0$. Panel (b) shows that the lifetime of the scarred eigenstate far exceeds that of the eigenstate next to the scar state in energy at large frequencies. Panels (c) and (d) show the spectral functions $f(0)$ of eigenstates of $H_0$ as a function of energy for $V$ and $[V, [H_0,V]]$ respectively. The scar state is an outlier with smaller values; the smaller $[V,[H_0,V]]$ spectral function in particular is responsible for the large-$\omega$ difference between the scar and an eigenstate of similar energy in (b).  } 
\label{fig:main3}
\end{figure}

The interzone and intrazone rates also accurately describe the fidelity decay of scarred eigenstates of $H_0$. Fig.~\ref{fig:main3}(b) compares the decay timescale for the scar state in the $S^z=0$ sector to an eigenstate with a similar energy. The scar is noticeably longer-lived, especially in the large frequency regime where the scar lifetime is an order of magnitude larger than that of the non-scarred eigenstate.

We interpret this behavior in terms of the spectral functions of $V$ and $[V,[H_0,V]]$, which set the interzone and intrazone rates respectively (Eqs.~\eqref{eq:interzone_sin} and \eqref{eq:intrazone_sin}).  In Fig.~\ref{fig:main3}(c) and (d), we plot these spectral functions evaluated at zero frequency and see that the scar state is an outlier relative to other states in the same symmetry sector - the smaller values of the spectral function lead to smaller decay rates. Note that $\mathcal{A}_\psi^{V}(0)$ does not enter the interzone rate directly except for setting an overall scale for the rapidly decaying $\mathcal{A}_\psi^{V}(\pm \omega)$, while $\mathcal{A}_\psi^{[V,[H_0,V]]}(0)$ enters the intrazone rate directly. The small value of $\mathcal{A}_\psi^{[V,[H_0,V]]}(0)$ in the scar state relative to neighboring eigenstates in energy explains the order-of-magnitude larger lifetime of the scarred state at large frequencies. 

The long lifetimes of scars at large frequencies are reflected in the dynamics of product states formed from superpositions of scar states. The product state $\bigotimes_{i=1}^L \frac{|+\rangle_i + (-1)^i |-\rangle_i}{\sqrt{2}}$, which is a superposition of scar states from different $S^z$ sectors, shows perfect oscillations of period $\frac{\pi}{h}$ under the dynamics of $H_0 + h S^z$. As seen in Fig.~\ref{fig:main3}(a), under the full dynamics of $H(t) + h S^z$, there are still long-lived oscillations for large driving frequencies despite the scar-breaking drive in $H(t)$. Note that the product state in Fig.~\ref{fig:main3}(a) decays faster than the scar state in the $S^z=0$ sector in Fig.~\ref{fig:main3}(b). The product state is a superposition of the scarred states in even $S^z$ sectors; as the size of the relevant matrix elements of $V$ and $[V,[H_0,V]]$ differ from sector to sector, we do not expect the same decay rate as in Fig.~\ref{fig:main3}(b).

\textbf{\textit{Discussion--}} We have analytically and numerically shown that the eigenstates of $H_0$ survive for long times under high frequency driving. In particular, the fidelity decay timescale $\dt$ grows exponentially with drive frequency in an wide frequency range before crossing over to a quartic dependence. The two regimes correspond to the respective dominance of interzone and intrazone channels for decay, with analytical expressions for the decay rates given for a sinusoidal drive in Eqs.~\eqref{eq:interzone_sin} and \eqref{eq:intrazone_sin} and for more general drives in the Supplementary Material~\cite{Supplementary}.

Our results hold for all eigenstates of $H_0$; for scar states with non-thermal local density matrices, local density matrices do not relax for long times. Thus classic signatures of quantum scarring, such as persistent oscillations in local densities, are robust to generic high frequency driving, which is promising for Floquet engineering approaches for generating scarred Hamiltonians. An interesting direction for future work is to determine the generality of Fig.~\ref{fig:main3}(b), that is, whether scarred states generically live longer. 

Finally, there is a close relationship between prethermalization at rapid drive frequencies $\omega$ and $U(1)$ prethermalization in the presence of a large magnetic field $h S^z$~\cite{Abanin-Huveneers2017_rigorous}.  We describe the similarities and differences to the driven case in the Supplementary Material~\cite{Supplementary} via analogy between the Floquet zones and magnetization sectors. This connection to $U(1)$ prethermalization is especially notable because it has been shown that a large number of scarred models can be derived starting from models with an underlying Lie algebra, with scars living in different $U(1)$ sectors~\cite{mark_unified_2020, odea_tunnels_2020, pakrouski_many-body_2020}. The energies of the scar states can be split to obtain ``towers" of scar states by adding a $U(1)$ field. The connection above means that a large magnitude for this $U(1)$ field can then be used to endow scars with a prethermal robustness to perturbations that break the $U(1)$ symmetry.

\noindent \emph{Acknowledgements--} We are grateful to  Sarang Gopalakrishnan, Wen Wei Ho, David Huse, Tomotaka Kuwahara, David Long, Alan Morningstar, Anatoli Polkovnikov and Marcos Rigol for useful discussions.  This work was supported by the US Department of Energy, Office of Science, Basic Energy Sciences, under Early Career Award Nos. DE-SC0021111 (N.O. and V.K.), and the NSF under NSF DMR-1752759 (A.C.) and NSF DMR-1928166 (F. B.).  V.K. also acknowledges support from the Alfred P. Sloan Foundation through a Sloan Research Fellowship and the Packard Foundation through a Packard Fellowship in Science and Engineering. Numerical simulations were performed on Stanford Research Computing Center's Sherlock cluster.  We acknowledge the hospitality of the Kavli Institute for Theoretical Physics at the University of California, Santa Barbara (supported by NSF Grant PHY-1748958).

\bibliography{main}

\begin{thebibliography}{29}%
\makeatletter
\providecommand \@ifxundefined [1]{%
 \@ifx{#1\undefined}
}%
\providecommand \@ifnum [1]{%
 \ifnum #1\expandafter \@firstoftwo
 \else \expandafter \@secondoftwo
 \fi
}%
\providecommand \@ifx [1]{%
 \ifx #1\expandafter \@firstoftwo
 \else \expandafter \@secondoftwo
 \fi
}%
\providecommand \natexlab [1]{#1}%
\providecommand \enquote  [1]{``#1''}%
\providecommand \bibnamefont  [1]{#1}%
\providecommand \bibfnamefont [1]{#1}%
\providecommand \citenamefont [1]{#1}%
\providecommand \href@noop [0]{\@secondoftwo}%
\providecommand \href [0]{\begingroup \@sanitize@url \@href}%
\providecommand \@href[1]{\@@startlink{#1}\@@href}%
\providecommand \@@href[1]{\endgroup#1\@@endlink}%
\providecommand \@sanitize@url [0]{\catcode `\\12\catcode `\$12\catcode `\&12\catcode `\#12\catcode `\^12\catcode `\_12\catcode `\%12\relax}%
\providecommand \@@startlink[1]{}%
\providecommand \@@endlink[0]{}%
\providecommand \url  [0]{\begingroup\@sanitize@url \@url }%
\providecommand \@url [1]{\endgroup\@href {#1}{\urlprefix }}%
\providecommand \urlprefix  [0]{URL }%
\providecommand \Eprint [0]{\href }%
\providecommand \doibase [0]{http://dx.doi.org/}%
\providecommand \selectlanguage [0]{\@gobble}%
\providecommand \bibinfo  [0]{\@secondoftwo}%
\providecommand \bibfield  [0]{\@secondoftwo}%
\providecommand \translation [1]{[#1]}%
\providecommand \BibitemOpen [0]{}%
\providecommand \bibitemStop [0]{}%
\providecommand \bibitemNoStop [0]{.\EOS\space}%
\providecommand \EOS [0]{\spacefactor3000\relax}%
\providecommand \BibitemShut  [1]{\csname bibitem#1\endcsname}%
\let\auto@bib@innerbib\@empty
\bibitem [{\citenamefont {Bukov}\ \emph {et~al.}(2015)\citenamefont {Bukov}, \citenamefont {D'Alessio},\ and\ \citenamefont {Polkovnikov}}]{Bukov:2015}%
  \BibitemOpen
  \bibfield  {author} {\bibinfo {author} {\bibfnamefont {Marin}\ \bibnamefont {Bukov}}, \bibinfo {author} {\bibfnamefont {Luca}\ \bibnamefont {D'Alessio}}, \ and\ \bibinfo {author} {\bibfnamefont {Anatoli}\ \bibnamefont {Polkovnikov}},\ }\bibfield  {title} {\enquote {\bibinfo {title} {Universal high-frequency behavior of periodically driven systems: from dynamical stabilization to floquet engineering},}\ }\href@noop {} {\bibfield  {journal} {\bibinfo  {journal} {Advances in Physics}\ }\textbf {\bibinfo {volume} {64}},\ \bibinfo {pages} {139--226} (\bibinfo {year} {2015})}\BibitemShut {NoStop}%
\bibitem [{\citenamefont {Eckardt}(2017)}]{Eckardt:2017}%
  \BibitemOpen
  \bibfield  {author} {\bibinfo {author} {\bibfnamefont {Andr\'e}\ \bibnamefont {Eckardt}},\ }\bibfield  {title} {\enquote {\bibinfo {title} {Colloquium: Atomic quantum gases in periodically driven optical lattices},}\ }\href {\doibase 10.1103/RevModPhys.89.011004} {\bibfield  {journal} {\bibinfo  {journal} {Rev. Mod. Phys.}\ }\textbf {\bibinfo {volume} {89}},\ \bibinfo {pages} {011004} (\bibinfo {year} {2017})}\BibitemShut {NoStop}%
\bibitem [{\citenamefont {Oka}\ and\ \citenamefont {Kitamura}(2019)}]{Oka:2019}%
  \BibitemOpen
  \bibfield  {author} {\bibinfo {author} {\bibfnamefont {Takashi}\ \bibnamefont {Oka}}\ and\ \bibinfo {author} {\bibfnamefont {Sota}\ \bibnamefont {Kitamura}},\ }\bibfield  {title} {\enquote {\bibinfo {title} {Floquet engineering of quantum materials},}\ }\href@noop {} {\bibfield  {journal} {\bibinfo  {journal} {Annual Review of Condensed Matter Physics}\ }\textbf {\bibinfo {volume} {10}},\ \bibinfo {pages} {387--408} (\bibinfo {year} {2019})}\BibitemShut {NoStop}%
\bibitem [{\citenamefont {Harper}\ \emph {et~al.}(2020)\citenamefont {Harper}, \citenamefont {Roy}, \citenamefont {Rudner},\ and\ \citenamefont {Sondhi}}]{Harper-Sondhi2020_review}%
  \BibitemOpen
  \bibfield  {author} {\bibinfo {author} {\bibfnamefont {Fenner}\ \bibnamefont {Harper}}, \bibinfo {author} {\bibfnamefont {Rahul}\ \bibnamefont {Roy}}, \bibinfo {author} {\bibfnamefont {Mark~S.}\ \bibnamefont {Rudner}}, \ and\ \bibinfo {author} {\bibfnamefont {S.L.}\ \bibnamefont {Sondhi}},\ }\bibfield  {title} {\enquote {\bibinfo {title} {Topology and broken symmetry in floquet systems},}\ }\href {\doibase 10.1146/annurev-conmatphys-031218-013721} {\bibfield  {journal} {\bibinfo  {journal} {Annual Review of Condensed Matter Physics}\ }\textbf {\bibinfo {volume} {11}},\ \bibinfo {pages} {345--368} (\bibinfo {year} {2020})}\BibitemShut {NoStop}%
\bibitem [{\citenamefont {Kitagawa}\ \emph {et~al.}(2010)\citenamefont {Kitagawa}, \citenamefont {Berg}, \citenamefont {Rudner},\ and\ \citenamefont {Demler}}]{Kitagawa-Demler_2010}%
  \BibitemOpen
  \bibfield  {author} {\bibinfo {author} {\bibfnamefont {Takuya}\ \bibnamefont {Kitagawa}}, \bibinfo {author} {\bibfnamefont {Erez}\ \bibnamefont {Berg}}, \bibinfo {author} {\bibfnamefont {Mark}\ \bibnamefont {Rudner}}, \ and\ \bibinfo {author} {\bibfnamefont {Eugene}\ \bibnamefont {Demler}},\ }\bibfield  {title} {\enquote {\bibinfo {title} {Topological characterization of periodically driven quantum systems},}\ }\href {\doibase 10.1103/PhysRevB.82.235114} {\bibfield  {journal} {\bibinfo  {journal} {Phys. Rev. B}\ }\textbf {\bibinfo {volume} {82}},\ \bibinfo {pages} {235114} (\bibinfo {year} {2010})}\BibitemShut {NoStop}%
\bibitem [{\citenamefont {Lindner}\ \emph {et~al.}(2011)\citenamefont {Lindner}, \citenamefont {Refael},\ and\ \citenamefont {Galitski}}]{lindner2011floquet}%
  \BibitemOpen
  \bibfield  {author} {\bibinfo {author} {\bibfnamefont {Netanel~H}\ \bibnamefont {Lindner}}, \bibinfo {author} {\bibfnamefont {Gil}\ \bibnamefont {Refael}}, \ and\ \bibinfo {author} {\bibfnamefont {Victor}\ \bibnamefont {Galitski}},\ }\bibfield  {title} {\enquote {\bibinfo {title} {Floquet topological insulator in semiconductor quantum wells},}\ }\href@noop {} {\bibfield  {journal} {\bibinfo  {journal} {Nature Physics}\ }\textbf {\bibinfo {volume} {7}},\ \bibinfo {pages} {490--495} (\bibinfo {year} {2011})}\BibitemShut {NoStop}%
\bibitem [{\citenamefont {Goldman}\ and\ \citenamefont {Dalibard}(2014)}]{Goldman:2014}%
  \BibitemOpen
  \bibfield  {author} {\bibinfo {author} {\bibfnamefont {N.}~\bibnamefont {Goldman}}\ and\ \bibinfo {author} {\bibfnamefont {J.}~\bibnamefont {Dalibard}},\ }\bibfield  {title} {\enquote {\bibinfo {title} {Periodically driven quantum systems: Effective hamiltonians and engineered gauge fields},}\ }\href {\doibase 10.1103/PhysRevX.4.031027} {\bibfield  {journal} {\bibinfo  {journal} {Phys. Rev. X}\ }\textbf {\bibinfo {volume} {4}},\ \bibinfo {pages} {031027} (\bibinfo {year} {2014})}\BibitemShut {NoStop}%
\bibitem [{\citenamefont {Khemani}\ \emph {et~al.}(2016)\citenamefont {Khemani}, \citenamefont {Lazarides}, \citenamefont {Moessner},\ and\ \citenamefont {Sondhi}}]{Khemani2016}%
  \BibitemOpen
  \bibfield  {author} {\bibinfo {author} {\bibfnamefont {V.}~\bibnamefont {Khemani}}, \bibinfo {author} {\bibfnamefont {A.}~\bibnamefont {Lazarides}}, \bibinfo {author} {\bibfnamefont {R.}~\bibnamefont {Moessner}}, \ and\ \bibinfo {author} {\bibfnamefont {S.~L.}\ \bibnamefont {Sondhi}},\ }\bibfield  {title} {\enquote {\bibinfo {title} {Phase structure of driven quantum systems},}\ }\href {\doibase 10.1103/PhysRevLett.116.250401} {\bibfield  {journal} {\bibinfo  {journal} {Phys. Rev. Lett.}\ }\textbf {\bibinfo {volume} {116}},\ \bibinfo {pages} {250401} (\bibinfo {year} {2016})}\BibitemShut {NoStop}%
\bibitem [{\citenamefont {Abanin}\ \emph {et~al.}(2015)\citenamefont {Abanin}, \citenamefont {De~Roeck},\ and\ \citenamefont {Huveneers}}]{Abanin-Huveneers2015_exponentially}%
  \BibitemOpen
  \bibfield  {author} {\bibinfo {author} {\bibfnamefont {Dmitry~A.}\ \bibnamefont {Abanin}}, \bibinfo {author} {\bibfnamefont {Wojciech}\ \bibnamefont {De~Roeck}}, \ and\ \bibinfo {author} {\bibfnamefont {Francois}\ \bibnamefont {Huveneers}},\ }\bibfield  {title} {\enquote {\bibinfo {title} {Exponentially slow heating in periodically driven many-body systems},}\ }\href {\doibase 10.1103/PhysRevLett.115.256803} {\bibfield  {journal} {\bibinfo  {journal} {Phys. Rev. Lett.}\ }\textbf {\bibinfo {volume} {115}},\ \bibinfo {pages} {256803} (\bibinfo {year} {2015})}\BibitemShut {NoStop}%
\bibitem [{\citenamefont {Mori}\ \emph {et~al.}(2016)\citenamefont {Mori}, \citenamefont {Kuwahara},\ and\ \citenamefont {Saito}}]{Mori-Saito2016_rigorous}%
  \BibitemOpen
  \bibfield  {author} {\bibinfo {author} {\bibfnamefont {Takashi}\ \bibnamefont {Mori}}, \bibinfo {author} {\bibfnamefont {Tomotaka}\ \bibnamefont {Kuwahara}}, \ and\ \bibinfo {author} {\bibfnamefont {Keiji}\ \bibnamefont {Saito}},\ }\bibfield  {title} {\enquote {\bibinfo {title} {Rigorous bound on energy absorption and generic relaxation in periodically driven quantum systems},}\ }\href {\doibase 10.1103/PhysRevLett.116.120401} {\bibfield  {journal} {\bibinfo  {journal} {Phys. Rev. Lett.}\ }\textbf {\bibinfo {volume} {116}},\ \bibinfo {pages} {120401} (\bibinfo {year} {2016})}\BibitemShut {NoStop}%
\bibitem [{\citenamefont {Kuwahara}\ \emph {et~al.}(2016)\citenamefont {Kuwahara}, \citenamefont {Mori},\ and\ \citenamefont {Saito}}]{Kuwahara-Saito2016_floquet}%
  \BibitemOpen
  \bibfield  {author} {\bibinfo {author} {\bibfnamefont {Tomotaka}\ \bibnamefont {Kuwahara}}, \bibinfo {author} {\bibfnamefont {Takashi}\ \bibnamefont {Mori}}, \ and\ \bibinfo {author} {\bibfnamefont {Keiji}\ \bibnamefont {Saito}},\ }\bibfield  {title} {\enquote {\bibinfo {title} {Floquet-magnus theory and generic transient dynamics in periodically driven many-body quantum systems},}\ }\href {\doibase https://doi.org/10.1016/j.aop.2016.01.012} {\bibfield  {journal} {\bibinfo  {journal} {Annals of Physics}\ }\textbf {\bibinfo {volume} {367}},\ \bibinfo {pages} {96--124} (\bibinfo {year} {2016})}\BibitemShut {NoStop}%
\bibitem [{\citenamefont {Abanin}\ \emph {et~al.}(2017)\citenamefont {Abanin}, \citenamefont {De~Roeck}, \citenamefont {Ho},\ and\ \citenamefont {Huveneers}}]{Abanin-Huveneers2017_rigorous}%
  \BibitemOpen
  \bibfield  {author} {\bibinfo {author} {\bibfnamefont {Dmitry}\ \bibnamefont {Abanin}}, \bibinfo {author} {\bibfnamefont {Wojciech}\ \bibnamefont {De~Roeck}}, \bibinfo {author} {\bibfnamefont {Wen~Wei}\ \bibnamefont {Ho}}, \ and\ \bibinfo {author} {\bibfnamefont {Fran{\c{c}}ois}\ \bibnamefont {Huveneers}},\ }\bibfield  {title} {\enquote {\bibinfo {title} {A rigorous theory of many-body prethermalization for periodically driven and closed quantum systems},}\ }\href {\doibase 10.1007/s00220-017-2930-x} {\bibfield  {journal} {\bibinfo  {journal} {Communications in Mathematical Physics}\ }\textbf {\bibinfo {volume} {354}},\ \bibinfo {pages} {809--827} (\bibinfo {year} {2017})}\BibitemShut {NoStop}%
\bibitem [{\citenamefont {Shiraishi}\ and\ \citenamefont {Mori}(2017)}]{shiraishi_systematic_2017}%
  \BibitemOpen
  \bibfield  {author} {\bibinfo {author} {\bibfnamefont {Naoto}\ \bibnamefont {Shiraishi}}\ and\ \bibinfo {author} {\bibfnamefont {Takashi}\ \bibnamefont {Mori}},\ }\bibfield  {title} {\enquote {\bibinfo {title} {Systematic {Construction} of {Counterexamples} to the {Eigenstate} {Thermalization} {Hypothesis}},}\ }\href {\doibase 10.1103/PhysRevLett.119.030601} {\bibfield  {journal} {\bibinfo  {journal} {Phys. Rev. Lett.}\ }\textbf {\bibinfo {volume} {119}},\ \bibinfo {pages} {030601} (\bibinfo {year} {2017})}\BibitemShut {NoStop}%
\bibitem [{\citenamefont {Bernien}\ \emph {et~al.}(2017)\citenamefont {Bernien}, \citenamefont {Schwartz}, \citenamefont {Keesling}, \citenamefont {Levine}, \citenamefont {Omran}, \citenamefont {Pichler}, \citenamefont {Choi}, \citenamefont {Zibrov}, \citenamefont {Endres}, \citenamefont {Greiner}, \citenamefont {Vuletić},\ and\ \citenamefont {Lukin}}]{bernien_probing_2017}%
  \BibitemOpen
  \bibfield  {author} {\bibinfo {author} {\bibfnamefont {Hannes}\ \bibnamefont {Bernien}}, \bibinfo {author} {\bibfnamefont {Sylvain}\ \bibnamefont {Schwartz}}, \bibinfo {author} {\bibfnamefont {Alexander}\ \bibnamefont {Keesling}}, \bibinfo {author} {\bibfnamefont {Harry}\ \bibnamefont {Levine}}, \bibinfo {author} {\bibfnamefont {Ahmed}\ \bibnamefont {Omran}}, \bibinfo {author} {\bibfnamefont {Hannes}\ \bibnamefont {Pichler}}, \bibinfo {author} {\bibfnamefont {Soonwon}\ \bibnamefont {Choi}}, \bibinfo {author} {\bibfnamefont {Alexander~S.}\ \bibnamefont {Zibrov}}, \bibinfo {author} {\bibfnamefont {Manuel}\ \bibnamefont {Endres}}, \bibinfo {author} {\bibfnamefont {Markus}\ \bibnamefont {Greiner}}, \bibinfo {author} {\bibfnamefont {Vladan}\ \bibnamefont {Vuletić}}, \ and\ \bibinfo {author} {\bibfnamefont {Mikhail~D.}\ \bibnamefont {Lukin}},\ }\bibfield  {title} {\enquote {\bibinfo {title} {Probing many-body dynamics on a 51-atom quantum simulator},}\ }\href {\doibase 10.1038/nature24622} {\bibfield  {journal}
  {\bibinfo  {journal} {Nature}\ }\textbf {\bibinfo {volume} {551}},\ \bibinfo {pages} {579--584} (\bibinfo {year} {2017})}\BibitemShut {NoStop}%
\bibitem [{\citenamefont {Serbyn}\ \emph {et~al.}(2021)\citenamefont {Serbyn}, \citenamefont {Abanin},\ and\ \citenamefont {Papi{\'c}}}]{Serbyn-Papic2021_review}%
  \BibitemOpen
  \bibfield  {author} {\bibinfo {author} {\bibfnamefont {Maksym}\ \bibnamefont {Serbyn}}, \bibinfo {author} {\bibfnamefont {Dmitry~A.}\ \bibnamefont {Abanin}}, \ and\ \bibinfo {author} {\bibfnamefont {Zlatko}\ \bibnamefont {Papi{\'c}}},\ }\bibfield  {title} {\enquote {\bibinfo {title} {Quantum many-body scars and weak breaking of ergodicity},}\ }\href {\doibase 10.1038/s41567-021-01230-2} {\bibfield  {journal} {\bibinfo  {journal} {Nature Physics}\ }\textbf {\bibinfo {volume} {17}},\ \bibinfo {pages} {675--685} (\bibinfo {year} {2021})}\BibitemShut {NoStop}%
\bibitem [{\citenamefont {Moudgalya}\ \emph {et~al.}(2022)\citenamefont {Moudgalya}, \citenamefont {Bernevig},\ and\ \citenamefont {Regnault}}]{Moudgalya-Regnault2021_review}%
  \BibitemOpen
  \bibfield  {author} {\bibinfo {author} {\bibfnamefont {Sanjay}\ \bibnamefont {Moudgalya}}, \bibinfo {author} {\bibfnamefont {B~Andrei}\ \bibnamefont {Bernevig}}, \ and\ \bibinfo {author} {\bibfnamefont {Nicolas}\ \bibnamefont {Regnault}},\ }\bibfield  {title} {\enquote {\bibinfo {title} {Quantum many-body scars and hilbert space fragmentation: a review of exact results},}\ }\href {\doibase 10.1088/1361-6633/ac73a0} {\bibfield  {journal} {\bibinfo  {journal} {Reports on Progress in Physics}\ }\textbf {\bibinfo {volume} {85}},\ \bibinfo {pages} {086501} (\bibinfo {year} {2022})}\BibitemShut {NoStop}%
\bibitem [{\citenamefont {Chandran}\ \emph {et~al.}(2023)\citenamefont {Chandran}, \citenamefont {Iadecola}, \citenamefont {Khemani},\ and\ \citenamefont {Moessner}}]{Chandran-Moessner2022_review}%
  \BibitemOpen
  \bibfield  {author} {\bibinfo {author} {\bibfnamefont {Anushya}\ \bibnamefont {Chandran}}, \bibinfo {author} {\bibfnamefont {Thomas}\ \bibnamefont {Iadecola}}, \bibinfo {author} {\bibfnamefont {Vedika}\ \bibnamefont {Khemani}}, \ and\ \bibinfo {author} {\bibfnamefont {Roderich}\ \bibnamefont {Moessner}},\ }\bibfield  {title} {\enquote {\bibinfo {title} {Quantum many-body scars: A quasiparticle perspective},}\ }\href {\doibase 10.1146/annurev-conmatphys-031620-101617} {\bibfield  {journal} {\bibinfo  {journal} {Annual Review of Condensed Matter Physics}\ }\textbf {\bibinfo {volume} {14}},\ \bibinfo {pages} {443--469} (\bibinfo {year} {2023})}\BibitemShut {NoStop}%
\bibitem [{\citenamefont {Bluvstein}\ \emph {et~al.}(2021)\citenamefont {Bluvstein}, \citenamefont {Omran}, \citenamefont {Levine}, \citenamefont {Keesling}, \citenamefont {Semeghini}, \citenamefont {Ebadi}, \citenamefont {Wang}, \citenamefont {Michailidis}, \citenamefont {Maskara}, \citenamefont {Ho}, \citenamefont {Choi}, \citenamefont {Serbyn}, \citenamefont {Greiner}, \citenamefont {Vuletić},\ and\ \citenamefont {Lukin}}]{bluvstein_controlling_2021}%
  \BibitemOpen
  \bibfield  {author} {\bibinfo {author} {\bibfnamefont {D.}~\bibnamefont {Bluvstein}}, \bibinfo {author} {\bibfnamefont {A.}~\bibnamefont {Omran}}, \bibinfo {author} {\bibfnamefont {H.}~\bibnamefont {Levine}}, \bibinfo {author} {\bibfnamefont {A.}~\bibnamefont {Keesling}}, \bibinfo {author} {\bibfnamefont {G.}~\bibnamefont {Semeghini}}, \bibinfo {author} {\bibfnamefont {S.}~\bibnamefont {Ebadi}}, \bibinfo {author} {\bibfnamefont {T.~T.}\ \bibnamefont {Wang}}, \bibinfo {author} {\bibfnamefont {A.~A.}\ \bibnamefont {Michailidis}}, \bibinfo {author} {\bibfnamefont {N.}~\bibnamefont {Maskara}}, \bibinfo {author} {\bibfnamefont {W.~W.}\ \bibnamefont {Ho}}, \bibinfo {author} {\bibfnamefont {S.}~\bibnamefont {Choi}}, \bibinfo {author} {\bibfnamefont {M.}~\bibnamefont {Serbyn}}, \bibinfo {author} {\bibfnamefont {M.}~\bibnamefont {Greiner}}, \bibinfo {author} {\bibfnamefont {V.}~\bibnamefont {Vuletić}}, \ and\ \bibinfo {author} {\bibfnamefont {M.~D.}\ \bibnamefont {Lukin}},\ }\bibfield  {title} {\enquote {\bibinfo
  {title} {Controlling quantum many-body dynamics in driven {Rydberg} atom arrays},}\ }\href {\doibase 10.1126/science.abg2530} {\bibfield  {journal} {\bibinfo  {journal} {Science}\ }\textbf {\bibinfo {volume} {371}},\ \bibinfo {pages} {1355--1359} (\bibinfo {year} {2021})}\BibitemShut {NoStop}%
\bibitem [{\citenamefont {Maskara}\ \emph {et~al.}(2021)\citenamefont {Maskara}, \citenamefont {Michailidis}, \citenamefont {Ho}, \citenamefont {Bluvstein}, \citenamefont {Choi}, \citenamefont {Lukin},\ and\ \citenamefont {Serbyn}}]{maskara_discrete_2021}%
  \BibitemOpen
  \bibfield  {author} {\bibinfo {author} {\bibfnamefont {Nishad}\ \bibnamefont {Maskara}}, \bibinfo {author} {\bibfnamefont {Alexios~A.}\ \bibnamefont {Michailidis}}, \bibinfo {author} {\bibfnamefont {Wen~Wei}\ \bibnamefont {Ho}}, \bibinfo {author} {\bibfnamefont {Dolev}\ \bibnamefont {Bluvstein}}, \bibinfo {author} {\bibfnamefont {Soonwon}\ \bibnamefont {Choi}}, \bibinfo {author} {\bibfnamefont {Mikhail~D.}\ \bibnamefont {Lukin}}, \ and\ \bibinfo {author} {\bibfnamefont {Maksym}\ \bibnamefont {Serbyn}},\ }\bibfield  {title} {\enquote {\bibinfo {title} {Discrete time-crystalline order enabled by quantum many-body scars: entanglement steering via periodic driving},}\ }\href {\doibase 10.1103/PhysRevLett.127.090602} {\bibfield  {journal} {\bibinfo  {journal} {Phys. Rev. Lett.}\ }\textbf {\bibinfo {volume} {127}},\ \bibinfo {pages} {090602} (\bibinfo {year} {2021})},\ \bibinfo {note} {arXiv: 2102.13160}\BibitemShut {NoStop}%
\bibitem [{\citenamefont {Mukherjee}\ \emph {et~al.}(2020{\natexlab{a}})\citenamefont {Mukherjee}, \citenamefont {Nandy}, \citenamefont {Sen}, \citenamefont {Sen},\ and\ \citenamefont {Sengupta}}]{Mukherjee_collapse:2020}%
  \BibitemOpen
  \bibfield  {author} {\bibinfo {author} {\bibfnamefont {Bhaskar}\ \bibnamefont {Mukherjee}}, \bibinfo {author} {\bibfnamefont {Sourav}\ \bibnamefont {Nandy}}, \bibinfo {author} {\bibfnamefont {Arnab}\ \bibnamefont {Sen}}, \bibinfo {author} {\bibfnamefont {Diptiman}\ \bibnamefont {Sen}}, \ and\ \bibinfo {author} {\bibfnamefont {K.}~\bibnamefont {Sengupta}},\ }\bibfield  {title} {\enquote {\bibinfo {title} {Collapse and revival of quantum many-body scars via floquet engineering},}\ }\href {\doibase 10.1103/PhysRevB.101.245107} {\bibfield  {journal} {\bibinfo  {journal} {Phys. Rev. B}\ }\textbf {\bibinfo {volume} {101}},\ \bibinfo {pages} {245107} (\bibinfo {year} {2020}{\natexlab{a}})}\BibitemShut {NoStop}%
\bibitem [{\citenamefont {Mukherjee}\ \emph {et~al.}(2020{\natexlab{b}})\citenamefont {Mukherjee}, \citenamefont {Sen}, \citenamefont {Sen},\ and\ \citenamefont {Sengupta}}]{Mukherjee_dynamics:2020}%
  \BibitemOpen
  \bibfield  {author} {\bibinfo {author} {\bibfnamefont {Bhaskar}\ \bibnamefont {Mukherjee}}, \bibinfo {author} {\bibfnamefont {Arnab}\ \bibnamefont {Sen}}, \bibinfo {author} {\bibfnamefont {Diptiman}\ \bibnamefont {Sen}}, \ and\ \bibinfo {author} {\bibfnamefont {K.}~\bibnamefont {Sengupta}},\ }\bibfield  {title} {\enquote {\bibinfo {title} {Dynamics of the vacuum state in a periodically driven rydberg chain},}\ }\href {\doibase 10.1103/PhysRevB.102.075123} {\bibfield  {journal} {\bibinfo  {journal} {Phys. Rev. B}\ }\textbf {\bibinfo {volume} {102}},\ \bibinfo {pages} {075123} (\bibinfo {year} {2020}{\natexlab{b}})}\BibitemShut {NoStop}%
\bibitem [{Sup()}]{Supplementary}%
  \BibitemOpen
  \href@noop {} {}\bibinfo {note} {See Supplementary Material for details on time-dependent perturbation theory and the Floquet-Magnus expansion, which includes Ref. \cite{Fox1976_critique}}\BibitemShut {NoStop}%
\bibitem [{\citenamefont {Gorin}\ \emph {et~al.}(2006)\citenamefont {Gorin}, \citenamefont {Prosen}, \citenamefont {Seligman},\ and\ \citenamefont {Žnidarič}}]{gorin_loschmidt_2006}%
  \BibitemOpen
  \bibfield  {author} {\bibinfo {author} {\bibfnamefont {Thomas}\ \bibnamefont {Gorin}}, \bibinfo {author} {\bibfnamefont {Tomaž}\ \bibnamefont {Prosen}}, \bibinfo {author} {\bibfnamefont {Thomas~H.}\ \bibnamefont {Seligman}}, \ and\ \bibinfo {author} {\bibfnamefont {Marko}\ \bibnamefont {Žnidarič}},\ }\bibfield  {title} {\enquote {\bibinfo {title} {Dynamics of loschmidt echoes and fidelity decay},}\ }\href {\doibase https://doi.org/10.1016/j.physrep.2006.09.003} {\bibfield  {journal} {\bibinfo  {journal} {Physics Reports}\ }\textbf {\bibinfo {volume} {435}},\ \bibinfo {pages} {33--156} (\bibinfo {year} {2006})}\BibitemShut {NoStop}%
\bibitem [{\citenamefont {Zarate-Herrada}\ \emph {et~al.}(2023)\citenamefont {Zarate-Herrada}, \citenamefont {Santos},\ and\ \citenamefont {Torres-Herrera}}]{zarate-herrada_generalized_2023}%
  \BibitemOpen
  \bibfield  {author} {\bibinfo {author} {\bibfnamefont {David~A.}\ \bibnamefont {Zarate-Herrada}}, \bibinfo {author} {\bibfnamefont {Lea~F.}\ \bibnamefont {Santos}}, \ and\ \bibinfo {author} {\bibfnamefont {E.~Jonathan}\ \bibnamefont {Torres-Herrera}},\ }\bibfield  {title} {\enquote {\bibinfo {title} {Generalized survival probability},}\ }\href {\doibase 10.3390/e25020205} {\bibfield  {journal} {\bibinfo  {journal} {Entropy}\ }\textbf {\bibinfo {volume} {25}} (\bibinfo {year} {2023}),\ 10.3390/e25020205}\BibitemShut {NoStop}%
\bibitem [{\citenamefont {Schecter}\ and\ \citenamefont {Iadecola}(2019)}]{schecter_weak_2019}%
  \BibitemOpen
  \bibfield  {author} {\bibinfo {author} {\bibfnamefont {Michael}\ \bibnamefont {Schecter}}\ and\ \bibinfo {author} {\bibfnamefont {Thomas}\ \bibnamefont {Iadecola}},\ }\bibfield  {title} {\enquote {\bibinfo {title} {Weak {Ergodicity} {Breaking} and {Quantum} {Many}-{Body} {Scars} in {Spin}-1 {X} {Y} {Magnets}},}\ }\href {\doibase 10.1103/PhysRevLett.123.147201} {\bibfield  {journal} {\bibinfo  {journal} {Phys. Rev. Lett.}\ }\textbf {\bibinfo {volume} {123}},\ \bibinfo {pages} {147201} (\bibinfo {year} {2019})}\BibitemShut {NoStop}%
\bibitem [{\citenamefont {Mark}\ \emph {et~al.}(2020)\citenamefont {Mark}, \citenamefont {Lin},\ and\ \citenamefont {Motrunich}}]{mark_unified_2020}%
  \BibitemOpen
  \bibfield  {author} {\bibinfo {author} {\bibfnamefont {Daniel~K.}\ \bibnamefont {Mark}}, \bibinfo {author} {\bibfnamefont {Cheng-Ju}\ \bibnamefont {Lin}}, \ and\ \bibinfo {author} {\bibfnamefont {Olexei~I.}\ \bibnamefont {Motrunich}},\ }\bibfield  {title} {\enquote {\bibinfo {title} {Unified structure for exact towers of scar states in the {Affleck}-{Kennedy}-{Lieb}-{Tasaki} and other models},}\ }\href {\doibase 10.1103/PhysRevB.101.195131} {\bibfield  {journal} {\bibinfo  {journal} {Phys. Rev. B}\ }\textbf {\bibinfo {volume} {101}},\ \bibinfo {pages} {195131} (\bibinfo {year} {2020})}\BibitemShut {NoStop}%
\bibitem [{\citenamefont {O'Dea}\ \emph {et~al.}(2020)\citenamefont {O'Dea}, \citenamefont {Burnell}, \citenamefont {Chandran},\ and\ \citenamefont {Khemani}}]{odea_tunnels_2020}%
  \BibitemOpen
  \bibfield  {author} {\bibinfo {author} {\bibfnamefont {Nicholas}\ \bibnamefont {O'Dea}}, \bibinfo {author} {\bibfnamefont {Fiona}\ \bibnamefont {Burnell}}, \bibinfo {author} {\bibfnamefont {Anushya}\ \bibnamefont {Chandran}}, \ and\ \bibinfo {author} {\bibfnamefont {Vedika}\ \bibnamefont {Khemani}},\ }\bibfield  {title} {\enquote {\bibinfo {title} {From tunnels to towers: {Quantum} scars from {Lie} algebras and q -deformed {Lie} algebras},}\ }\href {\doibase 10.1103/PhysRevResearch.2.043305} {\bibfield  {journal} {\bibinfo  {journal} {Phys. Rev. Research}\ }\textbf {\bibinfo {volume} {2}},\ \bibinfo {pages} {043305} (\bibinfo {year} {2020})}\BibitemShut {NoStop}%
\bibitem [{\citenamefont {Pakrouski}\ \emph {et~al.}(2020)\citenamefont {Pakrouski}, \citenamefont {Pallegar}, \citenamefont {Popov},\ and\ \citenamefont {Klebanov}}]{pakrouski_many-body_2020}%
  \BibitemOpen
  \bibfield  {author} {\bibinfo {author} {\bibfnamefont {K.}~\bibnamefont {Pakrouski}}, \bibinfo {author} {\bibfnamefont {P.N.}\ \bibnamefont {Pallegar}}, \bibinfo {author} {\bibfnamefont {F.K.}\ \bibnamefont {Popov}}, \ and\ \bibinfo {author} {\bibfnamefont {I.R.}\ \bibnamefont {Klebanov}},\ }\bibfield  {title} {\enquote {\bibinfo {title} {Many-{Body} {Scars} as a {Group} {Invariant} {Sector} of {Hilbert} {Space}},}\ }\href {\doibase 10.1103/PhysRevLett.125.230602} {\bibfield  {journal} {\bibinfo  {journal} {Phys. Rev. Lett.}\ }\textbf {\bibinfo {volume} {125}},\ \bibinfo {pages} {230602} (\bibinfo {year} {2020})}\BibitemShut {NoStop}%
\bibitem [{\citenamefont {Fox}(1976)}]{Fox1976_critique}%
  \BibitemOpen
  \bibfield  {author} {\bibinfo {author} {\bibfnamefont {Ronald~Forrest}\ \bibnamefont {Fox}},\ }\bibfield  {title} {\enquote {\bibinfo {title} {Critique of the generalized cumulant expansion method},}\ }\href {\doibase 10.1063/1.523041} {\bibfield  {journal} {\bibinfo  {journal} {Journal of Mathematical Physics}\ }\textbf {\bibinfo {volume} {17}},\ \bibinfo {pages} {1148--1153} (\bibinfo {year} {1976})}\BibitemShut {NoStop}%
\end{thebibliography}%


\begin{thebibliography}{4}%
\makeatletter
\providecommand \@ifxundefined [1]{%
 \@ifx{#1\undefined}
}%
\providecommand \@ifnum [1]{%
 \ifnum #1\expandafter \@firstoftwo
 \else \expandafter \@secondoftwo
 \fi
}%
\providecommand \@ifx [1]{%
 \ifx #1\expandafter \@firstoftwo
 \else \expandafter \@secondoftwo
 \fi
}%
\providecommand \natexlab [1]{#1}%
\providecommand \enquote  [1]{``#1''}%
\providecommand \bibnamefont  [1]{#1}%
\providecommand \bibfnamefont [1]{#1}%
\providecommand \citenamefont [1]{#1}%
\providecommand \href@noop [0]{\@secondoftwo}%
\providecommand \href [0]{\begingroup \@sanitize@url \@href}%
\providecommand \@href[1]{\@@startlink{#1}\@@href}%
\providecommand \@@href[1]{\endgroup#1\@@endlink}%
\providecommand \@sanitize@url [0]{\catcode `\\12\catcode `\$12\catcode `\&12\catcode `\#12\catcode `\^12\catcode `\_12\catcode `\%12\relax}%
\providecommand \@@startlink[1]{}%
\providecommand \@@endlink[0]{}%
\providecommand \url  [0]{\begingroup\@sanitize@url \@url }%
\providecommand \@url [1]{\endgroup\@href {#1}{\urlprefix }}%
\providecommand \urlprefix  [0]{URL }%
\providecommand \Eprint [0]{\href }%
\providecommand \doibase [0]{http://dx.doi.org/}%
\providecommand \selectlanguage [0]{\@gobble}%
\providecommand \bibinfo  [0]{\@secondoftwo}%
\providecommand \bibfield  [0]{\@secondoftwo}%
\providecommand \translation [1]{[#1]}%
\providecommand \BibitemOpen [0]{}%
\providecommand \bibitemStop [0]{}%
\providecommand \bibitemNoStop [0]{.\EOS\space}%
\providecommand \EOS [0]{\spacefactor3000\relax}%
\providecommand \BibitemShut  [1]{\csname bibitem#1\endcsname}%
\let\auto@bib@innerbib\@empty
\bibitem [{\citenamefont {Fox}(1976)}]{Fox1976_critique}%
  \BibitemOpen
  \bibfield  {author} {\bibinfo {author} {\bibfnamefont {Ronald~Forrest}\ \bibnamefont {Fox}},\ }\bibfield  {title} {\enquote {\bibinfo {title} {Critique of the generalized cumulant expansion method},}\ }\href {\doibase 10.1063/1.523041} {\bibfield  {journal} {\bibinfo  {journal} {Journal of Mathematical Physics}\ }\textbf {\bibinfo {volume} {17}},\ \bibinfo {pages} {1148--1153} (\bibinfo {year} {1976})}\BibitemShut {NoStop}%
\bibitem [{\citenamefont {Kuwahara}\ \emph {et~al.}(2016)\citenamefont {Kuwahara}, \citenamefont {Mori},\ and\ \citenamefont {Saito}}]{Kuwahara-Saito2016_floquet}%
  \BibitemOpen
  \bibfield  {author} {\bibinfo {author} {\bibfnamefont {Tomotaka}\ \bibnamefont {Kuwahara}}, \bibinfo {author} {\bibfnamefont {Takashi}\ \bibnamefont {Mori}}, \ and\ \bibinfo {author} {\bibfnamefont {Keiji}\ \bibnamefont {Saito}},\ }\bibfield  {title} {\enquote {\bibinfo {title} {Floquet-magnus theory and generic transient dynamics in periodically driven many-body quantum systems},}\ }\href {\doibase https://doi.org/10.1016/j.aop.2016.01.012} {\bibfield  {journal} {\bibinfo  {journal} {Annals of Physics}\ }\textbf {\bibinfo {volume} {367}},\ \bibinfo {pages} {96--124} (\bibinfo {year} {2016})}\BibitemShut {NoStop}%
\bibitem [{\citenamefont {Abanin}\ \emph {et~al.}(2015)\citenamefont {Abanin}, \citenamefont {De~Roeck},\ and\ \citenamefont {Huveneers}}]{Abanin-Huveneers2015_exponentially}%
  \BibitemOpen
  \bibfield  {author} {\bibinfo {author} {\bibfnamefont {Dmitry~A.}\ \bibnamefont {Abanin}}, \bibinfo {author} {\bibfnamefont {Wojciech}\ \bibnamefont {De~Roeck}}, \ and\ \bibinfo {author} {\bibfnamefont {Francois}\ \bibnamefont {Huveneers}},\ }\bibfield  {title} {\enquote {\bibinfo {title} {Exponentially slow heating in periodically driven many-body systems},}\ }\href {\doibase 10.1103/PhysRevLett.115.256803} {\bibfield  {journal} {\bibinfo  {journal} {Phys. Rev. Lett.}\ }\textbf {\bibinfo {volume} {115}},\ \bibinfo {pages} {256803} (\bibinfo {year} {2015})}\BibitemShut {NoStop}%
\bibitem [{\citenamefont {Abanin}\ \emph {et~al.}(2017)\citenamefont {Abanin}, \citenamefont {De~Roeck}, \citenamefont {Ho},\ and\ \citenamefont {Huveneers}}]{Abanin-Huveneers2017_rigorous}%
  \BibitemOpen
  \bibfield  {author} {\bibinfo {author} {\bibfnamefont {Dmitry}\ \bibnamefont {Abanin}}, \bibinfo {author} {\bibfnamefont {Wojciech}\ \bibnamefont {De~Roeck}}, \bibinfo {author} {\bibfnamefont {Wen~Wei}\ \bibnamefont {Ho}}, \ and\ \bibinfo {author} {\bibfnamefont {Fran{\c{c}}ois}\ \bibnamefont {Huveneers}},\ }\bibfield  {title} {\enquote {\bibinfo {title} {A rigorous theory of many-body prethermalization for periodically driven and closed quantum systems},}\ }\href {\doibase 10.1007/s00220-017-2930-x} {\bibfield  {journal} {\bibinfo  {journal} {Communications in Mathematical Physics}\ }\textbf {\bibinfo {volume} {354}},\ \bibinfo {pages} {809--827} (\bibinfo {year} {2017})}\BibitemShut {NoStop}%
\end{thebibliography}%

\end{document}


\title{Supplemental Material for: Prethermal stability of eigenstates under high frequency Floquet driving}

\author{Nicholas O'Dea}
\affiliation{Department of Physics, Stanford University, Stanford, CA 94305, USA}
\author{Fiona Burnell}
\affiliation{Department of Physics, University of Minnesota Twin Cities, MN 55455, USA}
\author{Anushya Chandran}
\affiliation{Department of Physics, Boston University, MA 02215, USA}
\author{Vedika Khemani}
\affiliation{Department of Physics, Stanford University, Stanford, CA 94305, USA}
\maketitle
\onecolumngrid
\renewcommand{\thefigure}{S\arabic{figure}}
\renewcommand{\theequation}{S\arabic{equation}}

In this supplemental material, we share some additional numerics exploring heating and reduced density matrices, and we explain the details of how we extracted timescales from fidelity curves. We then give a detailed discussion of time-dependent perturbation theory for the log of the fidelity, with explicit expressions for interzone and intrazone rates given in Eqs.~\ref{eq:FGR_rate} and ~\ref{Eq:gen_intrazone} and specifically for the box drive in ~Eqs. \ref{Eq:box_interzone} and \ref{Eq:box_intrazone}.

\section{Additional details on numerics}
\subsection{Heating versus fidelity}
In this section, we clarify how heating and fidelity curves are influenced by the initial state. In particular, we compare an eigenstate of $H_0$ to a product state that is not an eigenstate of $H_0$ but has the same energy density. 

\begin{figure}[h]
\includegraphics[width=.7\linewidth]{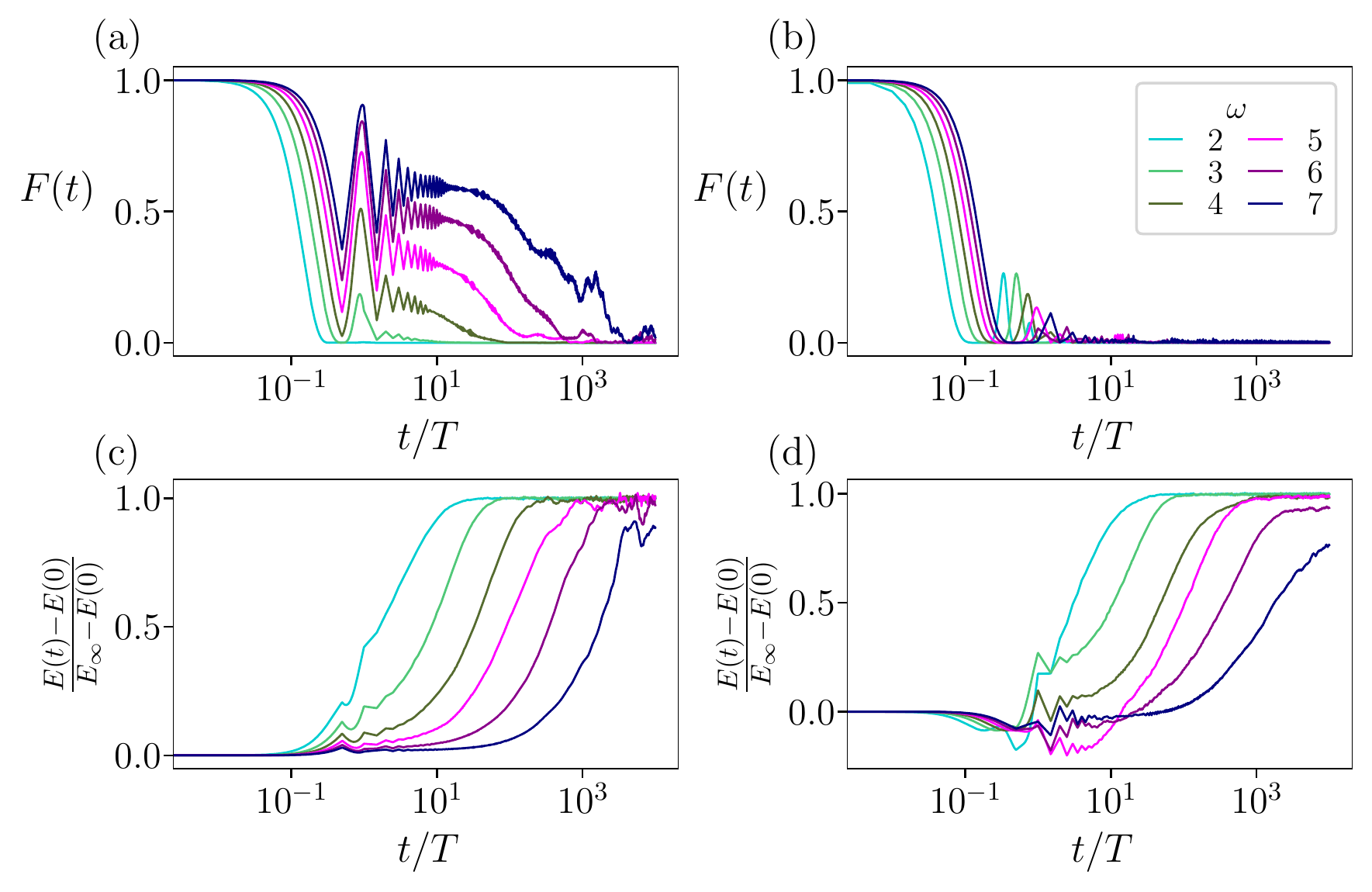}
\caption{Fidelity against time for (a) an eigenstate of $H_0$ and (b) a product state at the same energy density of $.56$. c) and d) show heating curves for the same eigenstate of $H_0$ and product state respectively. $L=14$, and the eigenstate is an energy density $.56$ eigenstate taken from the momentum $k=0$, magnetization $S^z=0$, inversion-symmetric, spin-flip-symmetric sector of $H_0$. The product state is at the same energy density and is proportional to $\bigotimes_{j=1}^L\left(\cos(\theta)|+\rangle_j + \sqrt{2} \sin(\theta)|0\rangle_j + \cos(\theta)|-\rangle_j\right)$ for $\theta=3\pi/8$.}
\label{fig:supp1}
\end{figure}

Proofs of prethermalization show that heating is slow regardless of the initial state: this is clear in Fig.~\ref{fig:supp1} c) and d), where both of these initial states show heating that is exponentially slow in frequency.

However, the results of our paper on long-lived fidelity only apply to all eigenstates of $H_0$. In particular, despite slow heating, the product state fidelity in Fig.~\ref{fig:supp1}b) rapidly decays in an $O(1)$ time. On the other hand, the eigenstate of $H_0$ indeed shows a long-lived fidelity in Fig.~\ref{fig:supp1}a).

\subsection{Timescale extraction}
In the main text, we extract timescales from fidelity curves against time, $F(t)$, and compare those timescales to Fermi's Golden Rule (FGR) predictions. In this appendix, we make our fitting procedure for extracting timescales from $F(t)$ explicit. 

\begin{figure}[h]
\includegraphics[width=\linewidth]{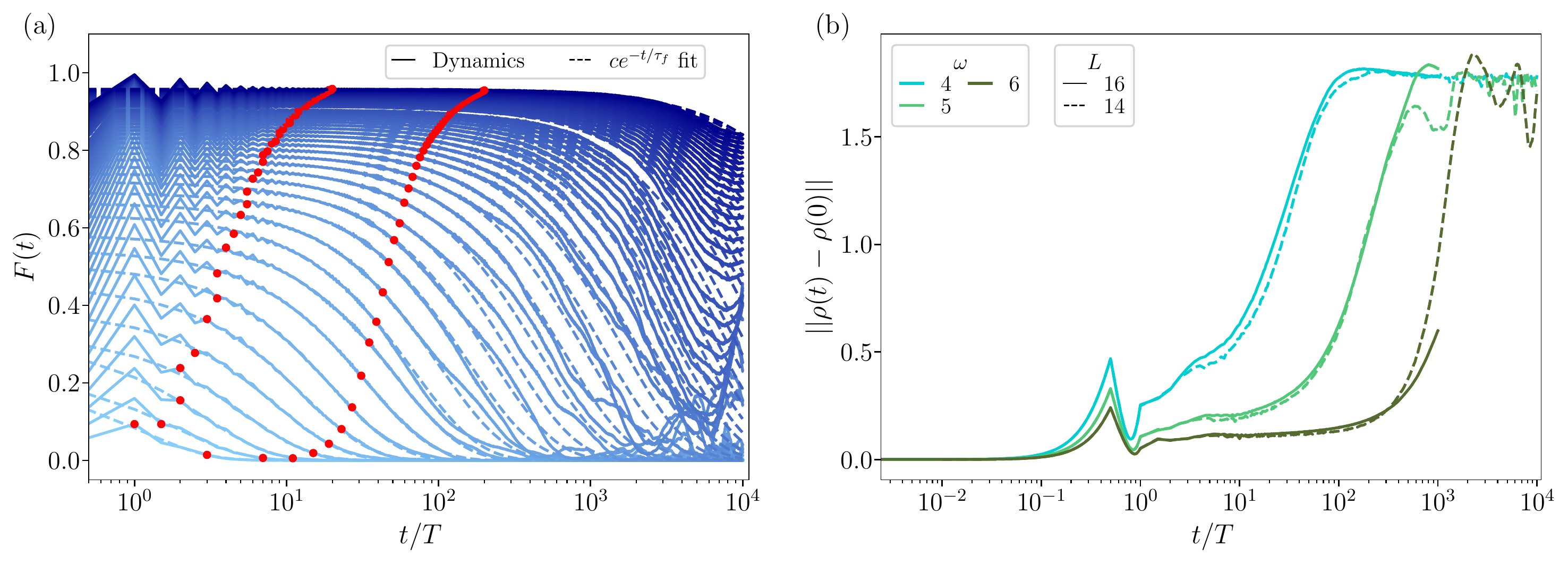}
\caption{a) Plotting the fidelity against time, $F(t)$, for an eigenstate of $H_0$ at zero energy density driven by $H(t)$ (Eq.~7), same as in Fig.~1a) in the main text. Here we show the curves for frequencies between $\omega=2.5$ and $14.75$ in steps of $.25$ from light to dark blue. Between each pair of red dots we fit a given $F(t)$ curve to the ansatz $F(t) = c e^{-t/\dt}$, and the resulting $\dt$ are used in Fig.~1b), ``Dynamics", in the main text. The dotted lines show the resulting $c e^{-t/\dt}$ curves. b) Trace-norm of the difference between $\rho(t)$ and $\rho(0)$ for the scar state in the $S^z=0$ sector of $H_0$ for system sizes $L=14,16$. }
\label{fig:supp_merged}
\end{figure}

We choose to fit our data to an ansatz of $F(t) = c e^{-t/\dt}$ for a limited range of times. Using a limited window to fit was motivated by the fact that FGR does not hold at early times and because our data may be susceptible to finite-size effects at very long times. We indeed begin to see oscillations in the fidelity at late times and large frequencies in Fig.~\ref{fig:supp_merged}a). The fitting windows for each $F(t)$ curve are explicitly shown in Fig.~\ref{fig:supp_merged}a), where we linearly interpolated between a window of $[1,3]$ and a window of $[20, 200]$ for frequencies between $\omega=2.5$ and $14.75$.

\subsection{Numerically estimating the FGR predictions}
In the main text, we compare extracted timescales (see above subsection) to predictions from Fermi's Golden Rule. In this subsection, we explain how we numerically calculate these analytical FGR predictions.

For the box drive considered in the main text, the FGR interzone and intrazone predictions for the fidelity decay rates of an eigenstate $|i\rangle$ of $H_0$ (derived in Section 2 of the supplementary material in Eqs.~\ref{Eq:box_interzone} and \ref{Eq:box_intrazone}) are 
\begin{equation}\label{Eq:box_interzone_temp}
R^{\text{inter}}_i = \frac{8 \lambda^2}{\pi} \sum_{n=0}^{\infty} \frac{\mathcal{A}_i^{V}((2n+1)\omega) + \mathcal{A}_i^{V}(-(2n+1)\omega)}{(2n+1)^2}
\end{equation}
\begin{equation}\label{Eq:box_intrazone_temp}
  R^{\text{intra}}_i    \approx  \frac{\pi^5\lambda^4}{288 \omega^4} \mathcal{A}_i^{[V,[H_0,V]]}(0)
\end{equation}
where the spectral functions are defined as 
\be
\mathcal{A}_i^{O}(\omega) =\sum_j  |O_{ij}|^2 \delta( (E_i - E_j) - \omega )
\ee
in the eigenbasis of $H_0$. 

We compute the spectral functions by diagonalizing $H_0$ and replacing the delta function $\delta( (E_i - E_j) - \omega)$ with a normalized indicator function $\frac{1}{W} \mathbbm{1}_{|(E_i-E_j)-\omega| \leq W/2}$ for a small energy window $W$. We also truncate the sum over harmonics in Eq.~\ref{Eq:box_interzone_temp}, which is valid because the spectral functions are rapdily decaying as a function of their arguments. In the main text, we use $W=.2$ and truncate the sum in Eq.~\ref{Eq:box_interzone_temp} to include the first six terms. 

\subsection{Local reduced density matrices}
In the main text, the primary quantity we probe is the many-body fidelity $F(t)$. As noted above, the fidelity decays with a rate that is extensive in system size. Here we consider the reduced density matrix on an $O(1)$ number of sites, $\rho(t)$, which should be asymptotically insensitive to system size at finite times.

In particular, in Fig.~\ref{fig:supp_merged}b), we calculate the change in the two-site reduced densitry matrix $\rho(t)$ relative to its initial condition $\rho(t=0)$ for the scar state in the $S^z=0$ sector of $H_0$. Specifically, we compute $||\rho(t)-\rho(0)||$ where $||A|| = Tr(\sqrt{A^\dagger A})$ is the trace-norm. This is a nice norm because it bounds the change in observables: $$Tr[(\rho(t) - \rho(0))O] \leq ||\rho(t) - \rho(0)|| \sigma_{\text{max}}(O)$$
where $\sigma_{\text{max}}(O)$ is the largest singular value of the operator $O$.
We see that the curves are not changing much as a function of system size, showing that our time-dependent results have largely converged at these times and frequencies.

Note that for non-scarred eigenstates of $H_0$ at similar energies to that of the scar, $\rho(0)$ already looks highly mixed, and so $\rho(t)$ is largely insensitive to the drive; i.e. $||\rho(t)-\rho(0)||$ is close to zero at all times for such states.

\section{Time-Dependent Perturbation theory}  \label{Sec:ManyDrives}

\subsection{Dyson series for the interaction picture $U_I(t)$}

We consider a Hamiltonian of the general form:
\begin{equation}
    H(t) = H_0 + \lambda g(t) V
\end{equation}

The time evolution by $H(t) = H_0 + V(t)$ is generated by $U(t) = U_I(t) U_0(t)$, where $U_0 = e^{-i H_0 t}$ and $U_I(t)$ is the unitary time-evolution operator in the interaction picture. $U_I(t)$ can be expanded in the Dyson series:
\begin{equation}\label{eq:suppDyson}
\begin{split}
    U_I(t)  &= \sum_{n=0}^\infty U^{(n)} \\
    U_{ij}^{(n)}&= (-i \lambda)^n \int^{t}_0 \int^{t_1}_0 ... \int^{t_{n-1}}_0 dt_1 ... dt_n \left( \prod_{j=1}^{n} f(t_j) \right ) \langle i | V(t_1)...V(t_n) |j \rangle
\end{split}
\end{equation}
where $U^{(0)} \equiv 1$, and $V(t) \equiv U_0^\dagger(t) V U_0(t)$.

To simplify the expression for $U_{ij}^{(n)}$, we express the periodic drive $g(t)$ as a Fourier series $g(t) =  \sum_{p} c_p e^{i p \omega t}$, where the fundamental frequency $\omega$ defines the period of the drive.   Inserting a complete set of eigenstates of $H_0$ between each power of $V$, we obtain:
\begin{equation} \label{Eq:UPert1}
U_{j_0 j_n}^{(n)}= (-i \lambda )^n  \sum_{p_1...p_n} c_{p_1}...c_{p_n} \sum_{j_1, j_2, ..., j_{n-1}}V_{j_0 j_1}V_{j_1 j_2}... V_{j_{n-1}j_n}  \int^{t}_0 \int^{t_1}_0 ... \int^{t_{n-1}}_0 dt_1 ... dt_n \prod_{q=1}^{n} e^{ i \Omega_q t_q }  
\end{equation}
where 
\be
\Omega_l = \Delta_{j_{l-1}  j_{l}} +p_l \omega   \ ,    \ \ \Delta_{ij} = \epsilon_{j} - \epsilon_i
\ee
with $\epsilon_j$ the unperturbed energy of eigenstate $|j \rangle$ of $H_0$.

To simplify this expression, we use the following identity:
\begin{equation} \label{Eq:PertSimplify}
\prod_{i=1}^{n} \left[ \int_0^{t_{i-1}} dt_i \right] \prod_{k=1}^{n} e^{i \Omega_{ k} t_k} = \frac{1}{i^{n}}\sum_{k=1}^n \frac{e^{i S_{0,k} t_0} -1}{\prod_{m=0; m\neq k }^n S_{m, k}  }
\end{equation}
where 
\begin{equation}
\begin{aligned} \label{Eq:Smk}
 S_{m, k}  = &\begin{cases}
 \sum_{l= m+1}^k \Omega_l
 & m<k \\
    - S_{ k,m} & m>k
\end{cases}
\end{aligned}
\end{equation} 
Inserting this in to Eq. (\ref{Eq:UPert1}), we obtain the general expression:
\begin{equation}
\begin{aligned}\label{eq:orderexpansion}
U_{j_0 j_n}^{(n)} &= (-\lambda)^n   \sum_{p_1...p_n} c_{p_1}...c_{p_n} \sum_{j_1, j_2, ..., j_{n-1}} V_{j_0 j_1}V_{j_1 j_2}... V_{j_{n-1}j_{n}} \sum_{k=1}^{n} \frac{e^{i S_{0 k} t} -1}{\prod_{m \neq k; m\geq 0}^{n} S_{m, k} } 
\end{aligned}
\end{equation}
which holds provided that $S_{mk} \neq 0$ for any pair $m,k$.  (We discuss the terms with $S_{mk} = 0$, known as secular terms, below).

To show that Eq. (\ref{Eq:PertSimplify})  holds for general $n$, we proceed by induction.
For $n=1$, the identity gives:
\begin{equation}
\int_0^{t_0 } dt_1  e^{i \Omega_{ 1} t_1} = \frac{1}{i}  \frac{e^{i  \Omega_{1} t_0} -1}{ \Omega_1}
\end{equation}
where we have used $S_{01} = \Omega_1$.  This is clearly true.
Therefore, let us assume that the formula holds for up to $n-1$ terms.  Then, since $\Omega_1 + S_{1k} = S_{0k}$:
\begin{equation}
\begin{split}
\prod_{i=1}^{n} \left[ \int_0^{t_{i-1}} dt_i \right] \prod_{k=1}^{n} e^{i \Omega_k t_k} 
=&  \frac{1}{i^{n-1}} \int_0^{t_0} dt_1  \left ( \sum_{k=2}^{n} \frac{e^{i S_{0 k} t_1}}{\prod_{m \neq k; m\geq 1}^{n} S_{m k}} -e^{i \Omega_1 t_1} \sum_{k=2}^{n} \frac{1}{\prod_{m \neq k; m\geq 1}^{n} S_{m k} } \right )
\end{split}
\end{equation}
We simplify the second term by noting that $ \sum_{k=1}^{n} \frac{1}{\prod_{m \neq k; m\geq 1}^{n} S_{m k} } = 0$. (This follows from the fact that, for any set of numbers $\{ x_j \}_{j=1}^n$, $\sum_j \prod_{k \neq j} \frac{1}{x_j - x_k} =0$. We take $x_j = \epsilon_j + \sum_{h = j+1}^n \Omega_h$, such that $x_j - x_k = S_{jk}$.) Thus we can simplify 
\begin{equation}
-e^{i \Omega_1 t_1} \sum_{k=2}^{n} \frac{1}{\prod_{m \neq k; m\geq 1}^{n} S_{m k} } = \frac{e^{i S_{01}t_1}}{\prod_{m \neq 1; m\geq1}^n S_{m 1}} \ .
\end{equation}
This gives the desired result:
\begin{equation}
\begin{split}
\prod_{i=1}^{n} \left[ \int_0^{t_{i-1}} dt_i \right] \prod_{k=1}^{n} e^{i \Omega_k t_k}  
= & \frac{1}{i^{n-1}} \int_0^{t_0} dt_1  \sum_{k=1}^{n} \frac{e^{i S_{0 k} t_1}}{\prod_{m \neq k; m\geq 1}^{n} S_{m k}} 
=  \frac{1}{i^{n}}\sum_{k=1}^{n} \frac{e^{i S_{0k} t_1} -1}{\prod_{m \neq k; m\geq 0}^{n} S_{m k}}.
\end{split}
\end{equation}

\subsection{Time dependence of fidelity via cumulant expansion}

Using the expression for the time evolution operator $U$ given by time-dependent perturbation theory, we now derive a perturbative expression for $\log (F (t))$, where 
\begin{equation} \label{Eq:LogF}
  \log (   F(t) ) = \log (   |\langle i|U|i\rangle|^2 )  = 2\Re [ \log((U_I)_{ii})]
\end{equation}
The $n^{th}$ order term in this series is simply the real part of the $n^{th}$ order term in the cumulant expansion of $U_{ii}$.  The general expressions for such cumulants can be found in \cite{Fox1976_critique}.

A property of some of the models of interest to us is that the Hamiltonian $H_0$ has a $U(1)$ symmetry, such that the total spin in the $z$ direction $S^z$ is conserved.  The perturbation $V = \sum_i S^x_i$, on the other hand, breaks this symmetry by changing the value of $S^z$ by exactly $\pm 1$.  It follows that for any eigenstate $|i \rangle $ of $H_0$, 
$\langle i | V^{2n +1} | i\rangle = 0$, since the states $|i \rangle$ and $V^{2n +1} | i\rangle $ cannot have the same value of $S^z$.  This means that there are no secular terms in $U^{(1)}$, and that $U^{(2n + 1)}_{ii} = 0$.
In the following discussions, we will assume the existence of such a symmetry, and drop any terms involving odd powers of $V$ from our calculation of $ \log (   F(t) ) $.  Note that this structure is \emph{not} essential to our main results; allowing $\langle i |V^{2n+1} |i \rangle \neq 0$  will only yield subleading-in-$\lambda$ corrections to the computed rates.

To fourth order, with $U^{(2n+1)}_{ii} = 0$, Eq. (\ref{Eq:LogF}) gives:
\begin{equation}
\begin{split}
\log     F(t) 
    &=2\Re\left(U_{ii}^{(2)} + (U_{ii}^{(4)} - \frac{1}{2} (U_{ii}^{(2)})^2) + O(\lambda^6)\right) \\
    &= 2\Re\left(U_{ii}^{(2)}\right) + (2\Re \left( U_{ii}^{(4)}\right) - \Re\left(U_{ii}^{(2)}\right)^2 +\Im\left(U_{ii}^{(2)}\right)^2 ) + O(\lambda^6)
\end{split}
\label{Eq:CumulatExpansion}
\end{equation}

We note that one could also consider a perturbative expansion for the fidelity itself, which takes the form:  
\begin{equation}
\begin{split}\label{eq:naivefidexpan}
    F(t) 
    &= (1 + U^{(2)}_{ii} + U^{(4)}_{ii} + ...)(1 + U^{(2)}_{ii} + U^{(4)}_{ii}+ ...)^* \\
    &= (1+(U^{(2)}_{ii} + U^{(2)*}_{ii})+(U^{(4)}_{ii} + U^{(2)}_{ii}U^{(2)*}_{ii} + U^{(4)*}_{ii}) )+ O(\lambda^6) \\ 
    &= 1+2\Re(U^{(2)}_{ii})+(2\Re(U^{(4)}_{ii}) +  |U^{(2)}_{ii}|^2) + O(\lambda^6)
\end{split}
\end{equation}
While the two expansions will match when all orders of perturbation theory are included, this form for $F(t)$ has two disadvantages.  First, at finite order in $\lambda$, it is not positive-definite, and indeed typically does fall below zero for some times when truncated at a low order in perturbation theory.  Second, a natural ansatz for the behavior of the fidelity is 
\begin{equation}
    F(t) = e^{-L^d f(\lambda, t)},
\end{equation}
which is expected because in a given time, there are $O(L^d)$ local changes in the initial state that can contribute to fidelity decay.  If this ansatz is correct, then the truncated cumulant expansion that describes $\log (F (t))$ should be proportional to $ L^d$ order by order in $\lambda$, while in the  expansion (\ref{eq:naivefidexpan}), multiple powers of $L^d$ appear at a given order of $\lambda$.

\subsection{$\log (F)$ to fourth order in $\lambda$}

Using the cumulant expansion (\ref{Eq:CumulatExpansion}), we can calculate the fourth-order correction to the fidelity.   Some care must be taken with secular terms.  Most of the terms in the series can be obtained from the fourth order terms in (\ref{eq:orderexpansion}) by Taylor expanding to leading order in $S_{02}$ and $S_{04}$ respectively.  The resulting secular terms grow linearly in time, and are purely imaginary.  At fourth order, however, additional secular terms are possible, and we must also account for the possibility that $S_{13}=0$ and $S_{02} =S_{24} = 0$.  These terms are not captured by the leading order Taylor expansion, but can be calculated directly from Eq. (\ref{Eq:UPert1}).

   For convenience, we will assume that all coefficients $c_{p_j}$ are either purely real, or purely imaginary, such that any product of an even number of these is real.  In this case, we find:
 \begin{eqnarray}
\begin{split} \label{Eq:CumulatExpansion2}
2\Re\left(U_{ii}^{(2)} \right ) 
    &=-4 \lambda^2 \sum_{j} |V_{i j}|^2 \left( \sum_{p_1 ,  p_2 } c_{p_1} c_{p_2}  \frac{\sin^2( S_{ij} t/2 ) }{S_{ij}(\Delta_{i j} - p_2 \omega)}  - \sum_{p_1 \neq -p_2 }  c_{p_1} c_{p_2}  \frac{\sin^2 (\omega( p_{1} + p_{2}) t/2) }{\omega( p_{1} + p_{2})(\Delta_{ij} - p_2 \omega)}  \right )  \\
    \end{split} \\ 
    \begin{split}
    2\Re \left( U_{ii}^{(4)}\right)  +  \Im\left(U_{ii}^{(2)}\right)^2  &=   2\lambda^4 \sum_{p_1,p_2, p_3, p_4} c_{p_1} c_{p_2} c_{p_3} c_{p_4}
     \sum_{j, k, l } V_{ij}V_{jk}V_{kl}V_{li} \left (  (1 - \delta_{ S_{ik} }) \left(  \frac{\sin^2 (S_{i k} t /2 )}{S_{i k} S'_{ki} S_{j k} S_{l k}}  \right)     \right . \\
     & \left . + ( 1 - \delta_{p_1 + p_2 + p_3 + p_4  } )
     \frac{\sin^2 (\omega
   ( p_{1}+p_{2}+p_{3}+p_{4})t/2) }{S'_{il}
   S'_{ik} S'_{ij} 
   \omega( p_{1}+p_{2}+p_{3}+p_{4})}
     \right . \\
     & \left . 
+
(1- \delta_{S_{jl}} ) \left (  \frac{\sin^2 ( S_{i j} t/2) }{S_{i j} S'_{ji} S_{l j} S_{k j}} + \frac{\sin^2 (  S_{i l} t/2 ) }{S_{i l} S'_{li}S_{jl} S_{k l}} \right  ) 
\right  ) \\ & 
 + \sum_{k, l} \delta_{p_2 + p_3} |V_{kl}|^2 |V_{il}|^2  \left(\frac{\sin^2 ( S_{il}t/2 ) (2 S_{kl}+ S_{il})}{S_{kl}^2
   S_{il}^3}-\frac{ 2 t \sin ( t  S_{il})}{S_{kl} S_{il} S'_{il} }\right)  \\ 
   & + \left (  \sum_{j} |V_{i j}|^2  \sum_{p_1 ,  p_2 } c_{p_1} c_{p_2}  \frac{\sin (S_{ij} t) }{S_{ij}(\Delta_{i j} - p_2 \omega)}  - \sum_{p_1 \neq -p_2 }  c_{p_1} c_{p_2}  \frac{\sin \omega( p_{1} + p_{2}) t }{\omega( p_{1} + p_{2})(\Delta_{ij} - \omega p_{2})} \right )^2  \\
    & - 2 \sum_{l, p_3 =  - p_4 }c_{p_3} c_{p_4}  \frac{ t } { \Delta_{il} - \omega p_4}  \sum_{j} |V_{i j}|^2  \left (  \sum_{p_1 ,  p_2 } c_{p_1} c_{p_2}  \frac{\sin (S_{ij} t) }{S_{ij}(\Delta_{i j} - \omega p_2)}  \right . \\
    & \left. - \sum_{p_1 \neq -p_2 }  c_{p_1} c_{p_2}  \frac{\sin \omega ( p_{1} + p_{2}) t }{\omega( p_{1} + p_{2})(\Delta_{ij} - \omega p_{2})} \right )
    \label{Eq:CumulatExpansion4}
\end{split} 
\end{eqnarray}
where, from Eq. (\ref{Eq:Smk})
\be
\begin{aligned}
S_{ij} = \Delta_{ij} + \omega p_1 \ , \ \ S_{ik} = \Delta_{ik} + \omega p_1  + \omega p_2 \ , \ \ S_{il} = \Delta_{il} + \omega p_1  + \omega p_2 + \omega p_3 \ , \ \ S_{jk} = \Delta_{jk} + \omega p_2 \ , \ \  S_{jl} = \Delta_{jl} + \omega p_2 + \omega p_3  \\
S_{kl} = \Delta_{kl} + \omega p_3 \ , \ \  S'_{ij} =- \Delta_{ij} + \omega p_2 + \omega p_3 + \omega p_4 \ , \ \  S'_{ik} =- \Delta_{ik} +  \omega p_3 + \omega p_4 \ , \ \ S'_{il} = -\Delta_{il} + \omega p_4 
\end{aligned} 
\ee
and as usual, $S_{ba} = - S_{ab}$, and similarly for $S'$.  
We note that though both $  2\Re \left( U_{ii}^{(4)}\right) $ and $  \Im\left(U_{ii}^{(2)}\right)^2$ have secular terms that grow with $t^2$, these cancel.  Of the remaining secular terms, which are linear in $t$, two are multiplied by a sum of incoherent sinusoids, and thus are not expected to diverge at long times as these incoherent sums fall off faster than $1/t$ for large $t$.  When $S_{ij}$ is very small, they can contribute to the late-time fidelity decay, which we discuss in the next subsection; however the corresponding matrix elements are always exponentially suppressed in $\omega/J$, and do not significantly impact the fourth-order fidelity decay.   

The final contribution, of the form
\be
2 t    \sum_{j} |V_{i j}|^2 \sum_{l, p_3  }c_{p_4} c_{-p_4}  \frac{1}{ (\Delta_{il} - \omega p_4)}  \sum_{p_1 \neq -p_2 }  c_{p_1} c_{p_2}  \frac{\sin \omega( p_{1} + p_{2}) t }{\omega( p_{1} + p_{2})(\Delta_{ij} - \omega p_{2}) } 
\ee
vanishes at integer multiples of the drive period.  

\subsection{Time scales in the perturbative expansion} \label{Sec:Freezeout}

We now turn to the time dependence of $\log(F)$, focusing on those terms that contribute significantly to fidelity decay at long times $t \gg \omega$.  We will show that these terms come in two types.  At second (and leading) order in $\lambda$, the contribution to the long-time fidelity decay comes only from inter-zone terms with exponentially suppressed matrix elements; these are the terms that contribute to heating.  At fourth order
in $\lambda$, there are inter-zone (heating) contributions, but also intra-zone processes, which do not contribute to heating but do contribute to fidelity decay.  In the next section, we will give explicit expressions for the contributions that both the second-order inter-zone and fourth order intra-zone terms make to the long-time fidelity decay rate.  

With the exception of the secular terms in the last line of Eq. (\ref{Eq:CumulatExpansion4}), which do not contribute to the long-time fidelity decay at $t = n T$, where $T$ is the drive period, we now show that  each term in the sums (\ref{Eq:CumulatExpansion2}-\ref{Eq:CumulatExpansion4}) will ``freeze out", in the sense that it no longer contributes to fidelity decay, after some time.

For each term in the sums, we can identify two distinct regimes.  The short-time regime describes times for which $S_{ab} t \ll 1$.  In this limit, we may Taylor expand the numerator to obtain:
\be
\begin{aligned} 
\frac{\sin(S_{a b } t)}{ S_{a b } }    \approx  t  \ , \ \ 
\frac{\sin^2 (S_{a b } t/2)}{ S_{a b } }   \approx  S_{ab} t^2/4  
\end{aligned} 
\ee
When $S_{ab} t \sim 1 $, this expansion breaks down, and we instead use the bounds:
\be
\sin^2 (S_{a b } t/2 ) \leq 1 \ , \ \ \ |\sin (S_{ab} t) | \leq 1
\ee
to upper-bound the contribution of each term to the log fidelity  by a constant. (The exception to this is the secular terms involving sums of products of $t  \sin S_{ab} t)$, for which this upper bound would yield a divergent result at large times, whereas the full sum is expected to be much smaller.)    This bound is attained for  $t = \pi / S_{a b}$ (or $t = \pi/(2 S_{ab})$, in the second case), after which the term in question has frozen out, in the sense that it ceases to contribute to the fidelity decay.

It is worth emphasizing that the overall growth of $|U_{ij}^{(n)}|$ can be much slower than these low-order expansions of individual terms suggest, as the leading-order contributions can cancel between different terms in the sum.  Such cancellations occur, for example, at very short times, when $\Omega_q t \ll 1$ for all $q$.   In this case the integrand in Eq. (\ref{Eq:UPert1}) is approximately constant, and $U^{(n)}_{ii} \sim \frac{(-i)^n}{n!}t^n (V^n)_{ii}$.  
In expressions of the form (\ref{eq:orderexpansion}), this results from the fact that when all complex exponentials are Taylor expanded, the coefficients of the first $n-1$ powers of $t$ vanish, leaving a leading-order dependence of $t^n$.

To understand the fidelity decay at time $t$, we  ignore those terms that have frozen out at shorter times, and consider a restricted sum over the remaining terms.  Up to fourth order, the long-time contributions at order $2n$ come from $\Re U^{(2n)}$, and are described by a restricted sum of the form (see Eq. (\ref{eq:orderexpansion}):
\begin{equation}
\begin{aligned}\label{eq:restrictedsum}
 \frac{t^2}{4} \sum_{p_1, p_2, ... p_{2n}} \delta(\sum_{i=1}^n p_i ) c_{p_1}c_{p_2}... c_{p_n} \sum_{k=1}^{2n} \left( \sum_{j_1, j_2, ..., j_{2n-1}|S_{j_0 j_k}(p) < 2/t } V_{j_0 j_1}V_{j_1 j_2}... V_{j_{2n-1}j_0} \frac{ S^2_{j_0 j_k} }{\prod_{m \neq k; m\geq 0}^{2n} S_{j_m j_k}} \right )
\end{aligned}
\end{equation}
Evidently, for $t$ large compared to the drive period, we count only terms where $S_{j_0 j_k} = \Delta_{j_0 j_k}$.  The number of states in the restricted sum is therefore given by
\begin{equation}
    N(t) = \int_{\epsilon_{j_0} - \frac{1}{t}}^{\epsilon_{j_0} + \frac{1}{t}} \rho(E) d E \sim  \frac{2}{t} \rho_0
\end{equation}
where in the final equality, we have assumed that the density of states is approximately constant, $\rho(E) \approx \rho_0$.  In other words, we obtain approximately $2/ (t \mathcal{J})$ of the total number of states in the sum, where $\mathcal{J}$ is the band-width  of $H_0$.  
At long times, we can roughly approximate the restricted sum to be $2/ (t \mathcal{J})$ of the unrestricted sum, giving: 
\begin{equation}
\begin{aligned}\label{eq:restrictedsum_approx}
\frac{t}{2 \mathcal{J} }\sum_{p_1, p_2, ... p_{2n}} \delta(\sum_{i=1}^n p_i ) c_{p_1}c_{p_2}... c_{p_n} \sum_{k=1}^{2n} \left( \sum_{j_1, j_2, ..., j_{2n-1} } V_{j_0 j_1}V_{j_1 j_2}... V_{j_{2n-1}j_0} \frac{ S^2_{j_0 j_k} }{\prod_{m \neq k; m\geq 0}^{2n} S_{j_m j_k} }\right )
\end{aligned}
\end{equation}
Since the term in the sum is time-independent, we therefore conclude that the log fidelity decay is approximately linear in time.  

In addition to terms of the form (\ref{eq:restrictedsum}), there are contributions from cross-terms in Eqs. (\ref{Eq:CumulatExpansion2} - \ref{Eq:CumulatExpansion4}), for which $S^2_{j_0 j_k}$ in the above expression is replaced by $S_{j_0 j_k} S_{j_0 j_l}$, and can appear with a different product of matrix elements.  These also lead to a fidelity decay that is at most linear in $t$, albeit with a different expression for the slope.  At fourth order, however, these all involve at least one high-frequency term, and thus do not contribute to the linear fidelity decay in the long-time limit.     

We now discuss the slope associated with Eq. (\ref{eq:restrictedsum}), to fourth order in perturbation theory.  
At second order, the fidelity decay (\ref{Eq:CumulatExpansion}) is described by $2\Re\left(U_{ii}^{(2)} \right ) $, given in Eq. (\ref{Eq:CumulatExpansion2}). 

This contains two types of terms.  First, for every choice of intermediate state $j$, the sum contains a term proportional to  $\sin^2( \omega( p_1 + p_2) t/2)$.  All of these terms freeze out in the sense of having attained their maximum impact on fidelity decay at a time-scale $ t \lessapprox \pi /( 2\omega)$.   However, because there are a large number of them, all of which are coherent, at time-scales beyond this, the fidelity shows evidence of oscillations at the frequency $\omega$.   (Note that the plots shown in the main text are stroboscopic after the first period, and these oscillations are not visible).

Second, there are terms proportional to $\sin^2( S_{ij} t/2 ) $, with $S_{i j} =   \Delta_{ij} + n \omega$, $ n \in \mathbb{Z}$.   
As described above, a useful upper bound for each of these terms is obtained by replacing
\begin{equation} 
\label{Eq:FreezeTime}
\sin ^2( S_{ij} t /2) \lessapprox 
\begin{cases}  
\left( S_{ij} t \right)^2 &   t < \frac{\pi}{S_{ij} }   \\ 
1 & \text{otherwise} \\
\end{cases}
\end{equation}
so that for $  t > \frac{\pi}{S_{ij} }$, the appropriate upper bound is constant in time, and the term has frozen out.  
 If $|\Delta_{ij}| <J$, the corresponding matrix elements are not suppressed in $\omega$, but for $\omega \gg J$ these terms freeze out at times  $t \sim  1/\omega$, and contribute to the fidelity decay only during the first few drive periods.  Such terms are a dominant source of the rapid fidelity decay at early times observed in the main text.    
The remaining terms at second order have $\Delta_{ij} + n \omega \ll \omega$.  These terms have matrix elements that are exponentially suppressed in $\omega/J$, 
and make a contribution to the fidelity decay that is linear in time, but with a slope that decreases as $e^{ - \omega / J}$.  
Thus at quadratic order,  the fidelity decay at times large compared to the drive period is expected to scale as: 
\be
\delta \log (F(t)) \sim - L^d t \lambda^2 e^{- \omega / J}
\ee

The quartic order  contribution to fidelity decay is given by $2 \Re U^{(4)}_{ii}  + (\Im U^{(2)}_{ii} )^2 -( \Re U^{(2)}_{ii} )^2$.  The contributions from  $(\Re U^{(2)}_{ii})^2$ either freeze out on the scale of a single drive period, or appear with matrix elements that are exponentially suppressed in $\omega/J$.  
The remaining fourth-order contribution is given in Eq. (\ref{Eq:CumulatExpansion4}), and contains both terms that freeze out on the scale of a single drive period (including a contribution from coherent oscillations at frequencies $n \omega$), and terms that contribute to the long-time fidelity decay which are multiplied by matrix elements suppressed by  $e^{- \omega /J}$.  
However, we also find a third type of term not present at second order.  Taking  $p_1 + p_2 = 0$, the first line of Eq. (\ref{Eq:CumulatExpansion4})  gives a contribution of the form:
\begin{equation}
\begin{split}
  -4 \sum_{p_1,p_3,p_4} c_{p_1} c_{-p_1} c_{p_3} c_{p_4} \sum_{j,k,l} V_{ij} V_{jk} V_{kl} V_{li} 
    \left(
   \frac{\sin^2 (t
   \Delta_{ik}/2)}{\Delta_{ik}(\Delta_{jk}-\omega p_{1})
   (\Delta_{kl}+\omega p_{3}) 
   (-\Delta_{ik}+\omega (p_{3}+p_{4}))} \right )
   \end{split} \ .
   \end{equation}
 For $\Delta_{ik} \lesssim J$ the corresponding matrix elements are not exponentially suppressed; thus we expect that these terms dominate the fidelity decay at times large compared to the drive period.

Thus, at long times, the dominant quartic-order contribution to the fidelity decay is described by the sum:
\begin{equation}
\begin{split} \label{Eq:LongTFidelity}
  & 4\lambda^4  \sum_{j,k,l} V_{ij} V_{jk} V_{kl} V_{li}    \frac{\sin^2 (t
   \Delta_{ik}/2)}{\Delta_{ik}}  \left [ 
    \sum_{p_1,p_3 }  \frac{1}{ \Delta_{ik}} 
   \frac{c_{p_1} c_{-p_1} c_{p_3} c_{-p_3}  }{(\Delta_{jk}-\omega p_{1})
   (\Delta_{kl}+\omega p_{3}) 
   } 
    \right .  \\
&   
\left .
 \sum_{p_1,p_3, p_4 \neq -p_3} \frac{c_{p_1} c_{-p_1} c_{p_3} c_{p_4}  }{(\Delta_{jk}-\omega p_{1})
   (\Delta_{kl}+\omega p_{3}) 
   (-\Delta_{ik}+ \omega p_{3}+ \omega p_4)}
  \right ] 
   \end{split} 
\end{equation}
where $c_{-p_j}$ is the coefficient associated with $-p_j$.  
The sum over frequencies in the first line gives:
\begin{equation} \label{Eq:cancellation}
\begin{split}
   &  \sum_{p_1,p_3 }   
   \frac{c_{p_1} c_{-p_1}c_{p_3} c_{-p_3}}{(\Delta_{jk}-\omega p_{1})
   (\Delta_{kl}+\omega p_{3}) 
   } =  \left(   \sum_{p_1> 0}  \frac{c_{p_1} c_{-p_1}}{(\Delta_{jk}-\omega p_1)} +  \frac{c_{p_1} c_{-p_1}}{(\Delta_{jk}+\omega p_1)}  \right )
   \left( \sum_{p_3>0 } \frac{c_{p_3} c_{-p_3}}{(\Delta_{kl}+\omega p_3) }+ \frac{c_{p_3} c_{-p_3}}{(\Delta_{kl}-\omega p_3) }\right ) \\
&   = \sum_{p_1> 0, p_3>0 } \frac{4c_{p_1} c_{-p_1} c_{p_3} c_{-p_3}  \Delta_{jk}\Delta_{kl}}{(\Delta_{jk}^2-(\omega p_1)^2) 
  (\Delta_{kl}^2-(\omega p_3)^2) } 
  \end{split} 
\end{equation}
If none of the matrix elements are exponentially suppressed, all energy differences must be small compared to $\omega$, and this term is of order $1/\omega^4$, rather than $1/\omega^2 \Delta_{ik}^2$ as one might naively have guessed.  
Similarly, summing over frequencies in the second line yields: 
\begin{equation}
\begin{split}
      \sum_{p_1,p_3, p_4 \neq - p_3}  \frac{c_{p_1} c_{-p_1} c_{p_3} c_{p_4}}{(\Delta_{jk}-\omega p_{1})
   (\Delta_{kl}+\omega p_{3}) 
   (-\Delta_{ik}+ \omega (p_{3} + p_4))}
= 
 \sum_{p_1 >0}  \frac{2c_{p_1} c_{-p_1} \Delta_{jk}}{(\Delta_{jk}^2-(\omega p_1)^2)}
  \sum_{p_3, p_4 \neq -p_3} \left( \frac{c_{p_3} c_{p_4}}{(\Delta_{kl}+\omega p_3) 
   (-\Delta_{ik}+ \omega (p_3 + p_4)} 
   \right )  \\
   \end{split}
\end{equation}
Again, when the matrix elements are not suppressed, this denominator is of order $\omega^4$.

Thus, at large $\omega/J$, the long-time fidelity decay is dominated by terms in  the first line Eq. (\ref{Eq:CumulatExpansion4}), which arise at fourth order in perturbation theory.  Their contribution to the fidelity decay is of the form
$
\delta \log (F(t) ) \sim  - t L^d \lambda^4/ \omega^4 $. These terms do not contribute to heating, as they are associated with processes that bring the system back to the original energy zone.  As we discuss in Sec. \ref{Sec:Magnus}, this scaling agrees with predictions of the Floquet Magnus expansion.

\section{Three FGR rates}

Here, we discuss how the three FGR rates presented in the main text are related to the perturbative expansions discussed above.  To simplify the expressions, we assume that the drive is either purely symmetric or purely anti-symmetric about $t=0$, such that its Fourier coefficients are either all real or all imaginary, and obey $c^*_{-p} = c_p$.  

\subsection{Heating rate to second order in $\lambda$}

First, to obtain a heating rate, 
we use the Dyson series (\ref{eq:suppDyson}) for $U_I(t)$ to perturbatively compute the time-dependent expectation of the energy $\langle E \rangle(t)$, where the expectation value is taken in a state that is initially an eigenstate $|i\rangle$ of $H_0$.  
Then
\begin{equation}
\begin{split}
    \langle E \rangle(t)
    &= \langle i| U_I^\dagger(t) H_0 U_I(t) |i \rangle \\
    &= \sum_{n=0} \sum_{k=0}^{n} \langle i| U^{(k)\dagger}(t) H_0 U^{(n-k)}(t) |i\rangle \equiv \sum_{n=0}  \langle E \rangle^{(n)}(t)
\end{split}
\end{equation}
To second order in $\lambda$, we have: 
\begin{equation}
\begin{split}
    \langle E \rangle^{(0)}(t) &= \langle i| H_0 |i\rangle = E_i  \ , \ \ \ 
    \langle E \rangle^{(1)}(t) = \langle i| U^{(1) \dagger}(t) H_0 + H_0 U^{(1)}(t) |i \rangle = 0 \\
    \langle E \rangle^{(2)}(t) &= \langle i|U^{(1) \dagger}(t)H_0U^{(1)}(t) + U^{(2)\dagger}(t)H_0 + H_0 U^{(2)}(t) |i \rangle   \\
    &= U^{(1) \dagger}_{ij}(t)  U^{(1)}_{ji}(t) E_j +2 \Re ( U^{(2)}(t))_{ii} E_i   \\
\end{split}
\end{equation}
Note that we have:
\be
\begin{aligned}
 ( U^{(1 \dag)})_{ij} U^{(1)}_{ji} &=    \lambda^2 \sum_{p_1 , p_2} c^*_{p_1} c_{p_2}   \sum_{j} | V_{i j} |^2  \left (\frac{e^{i (\Delta_{i j} - \omega p_{1}) t} -1}{(\Delta_{i j} - \omega p_{1})}  \right ) \left (\frac{e^{i (-\Delta_{i j} + \omega p_{2}) t} -1}{(-\Delta_{i j} + \omega p_{2})}  \right )  \n
&=  - 2 \lambda^2 \sum_{p_1 ,p_2} c^*_{p_1} c_{p_2}   \sum_{j} | V_{i j} |^2  \left (\frac{ \sin^2 ( \omega( p_{2} - p_{1}) t/2)  -  2 \sin^2 ( (\Delta_{i j} - \omega p_{1})t /2) 
 }{(\Delta_{i j} - \omega p_{1})(\Delta_{i j} - \omega p_{2})}  \right ) \ 
\end{aligned}
\ee
where in the last line, we have used the fact that $c^*_{p_1} c_{p_2} = c^*_{p_2} c_{p_1}$.  Using $c_{p_1} = c^*_{-p_1}$, we can re-write this as:
\be
 ( U^{(1 \dag)})_{ij} U^{(1)}_{ji}  = 4  \lambda^2 \sum_{p_1 ,p_2} c_{p_1} c_{p_2}   \sum_{j} | V_{i j} |^2  \left (\frac{ \sin^2 (S_{ij} t /2) - \frac{1}{2}  \sin^2 ( \omega( p_{2} + p_{1}) t/2)   
 }{S_{ij} (\Delta_{i j} - \omega p_{2})}  \right ) \ 
\ee
where as above, $S_{ij} =  \Delta_{i j} + \omega p_{1}$. We can simplify the second term in the parentheses by reindexing the sums $p_1\to -p_1, p_2 \to -p_2$ and using our simplifying assumption that $c_{p_1} c_{p_2} = c_{-p_1} c_{-p_2}$.
\be
\begin{split}
\sum_{ p_1 \neq -p_2 }  c_{p_1} c_{p_2}  \frac{\sin^2 (\omega( p_{1} + p_{2}) t/2) }{\omega( p_{1} + p_{2})(\Delta_{ij} - \omega p_{2})}  & = \frac{1}{2} \sum_{p_1 \neq -p_2 }\frac{\sin^2 ( \omega(p_{1} + p_{2}) t/2) }{ \omega (p_{1} + p_{2})} \left (\frac{ c_{p_1} c_{p_2}  }{\Delta_{ij} - \omega p_{2}}  -   \frac{c_{-p_1} c_{- p_2}  }{\Delta_{ij} +  \omega p_{1}}\right )  \\
&= \frac{1}{2} \sum_{p_1 \neq -p_2 } c_{p_1} c_{p_2}  \frac{\sin^2 (\omega( p_{1} + p_{2}) t/2) }{S_{ij} (\Delta_{ij} - \omega p_{2})}  \\
\end{split} 
\ee
Thus, we can simplify expression for $\langle E \rangle^{(2)}(t)$ to obtain:
\begin{equation}
\begin{split}
   \langle E \rangle^{(2)}(t) = \lambda^2 \sum_{j}  |V_{i j}|^2   \sum_{p_1,  p_2 } c_{p_1} c_{p_2}  
 \left( \frac{2 \sin^2 (\omega( p_{1} + p_{2}) t/2)  -4 \sin^2( S_{ij} t/2 ) }{S_{ij}(\Delta_{i j} - \omega p_2)} \right ) ( E_i - E_j)
\end{split}
\end{equation}

We can extract a single rate from this using the approximations in Fermi's Golden Rule.  First, we neglect terms that are rapidly oscillating in time.  This leaves only the second term in the parentheses, and only those terms with $S_{ij}$ very small.  Second, at long times, the dominant terms will be those for which both $S_{ij}$ and $\Delta_{ij} - \omega p_2$ are close to vanishing, i.e. terms for which $p_2 = - p_1$ (and hence $\Delta_{ij} - \omega p_2 = S_{ij}$), and $S_{ij} \equiv \Delta_{ij} + \omega p_1 \approx 0$.    For these terms, we treat the sincs as approximate delta functions, $\left( \frac{\sin(S_{ij} t /2 )}{S_{ij}} \right)^2 \sim   \frac{t  \pi}{2} \delta(S_{ij})$, to obtain the long-time heating rate: 
\begin{equation}\label{Eq:gen_heat}
    R^{\text{heat}}_i = \frac{d}{dt} \langle E \rangle^{(2)}(t)\big|_{\text{FGR}} = 2\pi \lambda^2 \sum_{p} c_{p} c_{- p}\omega p  \mathcal{A}_i^{V}(\omega p) 
\end{equation}
where 
\be
\mathcal{A}_i^{V}(\omega) =\sum_j  |V_{ij}|^2 \delta( \Delta_{ij} - \omega )
\ee
 is the spectral function for the operator $V$ at frequency $\omega p$,
and the symbol $|_{\text{FGR}}$ means under the assumptions of Fermi's Golden Rule. 

For example, for a sine drive of $g(t) = \sin(t)$, this reduces to 
\be
R^{\text{heat}}_i = \frac{\pi \lambda^2}{2} \omega (\mathcal{A}_i^{V}(\omega)-\mathcal{A}_i^{V}(-\omega))
\ee
For the case of the odd box-drive considered in the text, there are only odd-harmonics, and $c_{2n+1} c_{-(2n+1)} = \left(\frac{2}{\pi}\right)^2 \frac{1}{(2n+1)^2}$, yielding
\be
R^{\text{heat}}_i = \frac{8 \lambda^2}{\pi} \omega \sum_{n=0}^{\infty} \frac{\mathcal{A}_i^{V}((2n+1)\omega) - \mathcal{A}_i^{V}(-(2n+1)\omega)}{(2n+1)}
\ee
In practice, since $\mathcal{A}^V_i$ decays exponentially in its argument, the matrix elements associated with higher harmonics are strongly suppressed, and the sums are well-described at large frequencies by only including a few harmonics.

\subsection{Fidelity decay rate to second order in $\lambda$ } 
The same approach can be used to compute a fidelity decay rate through Fermi's golden rule.  To quadratic order in $\lambda$, $\log F$ is given by
$2\Re\left(U_{ii}^{(2)} \right ) $, and the terms relevant to our rate calculation are exactly as in the heating rate calculation above:  only those with $p_2 = - p_1$ and $\Delta_{ij} + \omega p_1 \approx 0$ enter into the Fermi's golden rule rate.  Letting $G^{(j)}$ be the $j^{th}$ order contribution to $\log F$, we thus obtain:
\begin{equation}\label{eq:FGR_rate}
    R^{\text{inter}}_i =  \frac{d}{dt} G^{(2)}(t)\big|_{\text{FGR}} = 2 \pi \lambda^2 \sum_{p_1} c_{p_1} c_{- p_1}  \mathcal{A}_i^{V}(\omega p_1)  \ .
\end{equation}
For a sine drive, this reduces to 
\begin{equation}\label{Eq:sin_interzone}
    R^{\text{inter}}_i = \frac{\pi \lambda^2}{2}  \left(\mathcal{A}_i^{V}(\omega) + \mathcal{A}_i^{V}(-\omega) \right)  \ .
\end{equation}
As noted above, for the case of the odd box-drive considered in the text, there are only odd-harmonics, and $c_{2n+1} c_{-(2n+1)} = \left(\frac{2}{\pi}\right)^2 \frac{1}{(2n+1)^2}$; this yields
\begin{equation}\label{Eq:box_interzone}
R^{\text{inter}}_i = \frac{8 \lambda^2}{\pi} \sum_{n=0}^{\infty} \frac{\mathcal{A}_i^{V}((2n+1)\omega) + \mathcal{A}_i^{V}(-(2n+1)\omega)}{(2n+1)^2}
\end{equation}
We identify these rates as stemming from interzone processes given that they have a similar form to the lowest order heating rate above and involve the same matrix elements.

\subsection{Fidelity decay rate to fourth order in $\lambda$ }
As discussed above, at fourth order, the most interesting terms that contribute to the long-time limit are those that do not contribute to heating, and are not suppressed exponentially in $\omega/J$.  These terms are given in Eq. (\ref{Eq:LongTFidelity}).    Of the terms listed there, only those in the first line, with $p_3 = - p_4$, contribute to the FGR rate.  
 Using the simplification in Eq. (\ref{Eq:simple28}), and the fact that $c_{p_j} c_{- p_j}$ is real, these can be expressed: 
 \begin{equation}
\begin{split}
  4\lambda^4  \sum_{k}  \frac{\sin^2 (t
   \Delta_{ik}/2)}{\Delta_{ik}^2}   \sum_{j,l} V_{ij} V_{jk} V_{il}^*  V_{lk}^* 
    \sum_{p_1>0,p_3>0} 
   \frac{4 c_{p_1} c_{-p_1} c_{p_3} c_{-p_3} \Delta_{jk} \Delta_{kl}}{(\Delta_{jk}^2-(\omega p_{1})^2)
   (\Delta_{kl}^2-(\omega p_{3})^2) } \\
   =  4 \lambda^4  \sum_{k}  \frac{\sin^2 (t
   \Delta_{ik}/2)}{\Delta_{ik}^2}  \left |  \sum_{j, p_1>0}  c_{p_1} c_{-p_1} V_{ij} V_{jk}     \frac{ 2 \Delta_{jk}}{\Delta_{jk}^2-(\omega p_1)^2 } \right |^2
   \end{split}  \ .
   \end{equation}
   As above, to obtain a Fermi's Golden rule rate we keep only those terms that do not freeze out until the longest times, and use $\text{lim}_{\Delta_{ik} \rightarrow 0}\left( \frac{ \sin (\Delta_{ik} t/2)}{ \Delta_{ik} } \right )^2 = \frac{ t \pi}{2} \delta( \Delta_{ik} )$.  
For large $\omega$, the matrix elements are strongly suppressed unless $\Delta_{jk} \ll \omega$.  Thus up to corrections of order $(\Delta_{jk}/\omega p_1)^2$, we have: 
\begin{equation}
 R^{\text{intra}}_i =\frac{d}{dt} G^{(4)}(t)\big|_{\text{FGR}}    \approx  
2\pi  \lambda^4 \bigg|\sum_{ p_1>0}   \frac{c_{p_1} c_{-p_1}}{(\omega p_1)^2} \bigg|^2 \sum_k  \delta( \Delta_{ik})   \left |  \sum_{j} 2  V_{ij} V_{jk}   \Delta_{jk} \right |^2
\end{equation}
When $\Delta_{ik}$ is close to zero, $2\Delta_{jk} \approx \Delta_{jk} - \Delta_{ij}$, and $\sum_{j} 2 V_{ij}V_{jk}\Delta_{jk} \approx [V,[H_0,V]]_{ik}$.  Thus, we obtain: 
\begin{equation}\label{Eq:gen_intrazone}
   R^{\text{intra}}_i    \approx   2\pi  \lambda^4 \bigg|\sum_{ p_1>0}   \frac{c_{p_1} c_{-p_1}}{(\omega p_1)^2}\bigg|^2\sum_{k}  \delta( \Delta _{ik})  \left| \langle i | [V,[H_0,V]] | k\rangle   \right|^2 = \mathcal{A}_i^{[V,[H_0,V]]}(0) \frac{2\pi \lambda^4}{\omega^4} \bigg| \sum_{ p_1>0}   \frac{c_{p_1} c_{-p_1}}{( p_1)^2}\bigg|^2    
\end{equation}
For a sine drive $g(t)=\sin(t)$, the expression simplifies to 
\begin{equation}\label{Eq:sin_intrazone}
  R^{\text{intra}}_i    \approx  \frac{\pi \lambda^4}{8 \omega^4} \mathcal{A}_i^{[V,[H_0,V]]}(0)
\end{equation}
For the odd box-drive considered in the text, the expression reduces to
\begin{equation}\label{Eq:box_intrazone}
  R^{\text{intra}}_i    \approx  \frac{\pi^5\lambda^4}{288 \omega^4} \mathcal{A}_i^{[V,[H_0,V]]}(0)
\end{equation}

\subsection{Comparison to Floquet-Magnus fidelity decay rate} \label{Sec:Magnus} 

As one might expect, the dominant fourth-order contribution to the FGR rate matches the rate predicted from a Floquet-Magnus/van Vleck expansion.  We now show this explicitly for the odd box drive used in our numerics.    

Before discussing the odd box drive, we discuss the even box drive, for which our expressions are somewhat simpler.  For an even box drive, $V(t) = V \text{sgn}(\cos(\omega t))$,  the Floquet-Magnus expansion up to second order (see, e.g.~\cite{Kuwahara-Saito2016_floquet}) gives the effective Hamiltonian: 
\begin{equation}
H_F \sim H_0 - \frac{\pi^2}{24}\frac{\lambda^2}{\omega^2} ([V,[H_0,V]]-[H_0,[H_0,V]]
\end{equation}

From the FGR expression (\ref{eq:FGR_rate}) for $G^{(2)}(t)$ for a constant perturbation, this yields a fidelity decay rate of 
\begin{equation}
    R^{\text{intra}}_i = \frac{\pi^5 \lambda^4}{288 \omega^4} \mathcal{A}_i^{[V,[H_0,V]]}(0)
\end{equation}
Note that the $[H_0,[H_0,V]]$ term drops out, since its matrix elements in the $H_0$ eigenbasis are $(E_i-E_j)^2V_{ij}$, which vanish for $E_i - E_j = 0$. As a technical note, $[V,[H_0,V]]$ can have a diagonal piece, which we absorb into $H_0$, shifting its energies, before running perturbation theory.  

We now turn to the odd box drive, $V(t) = V \text{sgn}(\sin(\omega t))$.  Physically, it is clear that this should give the same lowest-order rate, though we will see that the derivation is significantly more involved. 
For an odd drive, the Floquet Magnus expansion up to second order gives the effective Hamiltonian: 
\begin{equation}
H_F \sim H_0 - i\frac{\pi}{2} \frac{\lambda}{\omega} [H_0,V] + \frac{\pi^2}{6}\frac{\lambda^2}{\omega^2} [V,[H_0,V]]
\end{equation}
Analogously to $[H_0,[H_0,V]]$ above, the $[H_0,V]$ term drops out of the FGR expression for $G^{(2)}(t)$, yielding a contribution of
\begin{equation}
     \frac{d}{dt} G^{(2)}(t)\big|_{\text{FGR}}= \frac{\pi^5 \lambda^4}{18 \omega^4} \mathcal{A}_i^{[V,[H_0,V]]}(0)
\end{equation}

However, because $[H_0,V]$ is lower-order in $\lambda$ and $\omega$ relative to the $[V,[H_0,V]]$ term, fourth-order TDPT for a quench causes it to enter on the same footing as second order TDPT on $[V,[H_0,V]]$.

For a quench, fourth-order TDPT for a perturbation $\frac{\pi}{2} \frac{\lambda}{\omega} [H_0,V]$ yields (note that $[H_0,V]_{ij} = (E_i - E_j) V_{ij} = \Delta_{ij} V_{ij}$)
\begin{equation}
\begin{split}
    \Re\left( \langle i |U^{(4)}_I |i \rangle\right) &= \frac{\pi^4 \lambda^4}{8 \omega^4} \sum_{jkl}V_{ij}V_{jk}V_{kl}V_{li} \Delta_{ij}\Delta_{jk}\Delta_{kl}\Delta_{li} \left( \frac{\sin^2 (\Delta_{i j} t /2 )}{\Delta_{ij}^2 \Delta_{kj}\Delta_{lj}}+\frac{\sin^2 (\Delta_{i k} t /2 )}{\Delta_{ik}^2 \Delta_{jk}\Delta_{lk}}+\frac{\sin^2 (\Delta_{i l} t /2 )}{\Delta_{il}^2 \Delta_{jl}\Delta_{kl}} \right)
\end{split}
\end{equation}

The first and third terms in the parentheses contribute to the late-time behavior when $\Delta_{ij}$ and $\Delta_{il}$ are close to zero; however, these contributions are completely suppressed from the factors of $\Delta_{ij}$ and $\Delta_{li}$ from the matrix elements of $[H_0, V]$ appearing in the pre-factor.
The dominant term at late times is thus proportional to 
\begin{equation}
\sum_{jkl}V_{ij}V_{jk}V_{kl}V_{li} \frac{\sin^2 (\Delta_{i k} t /2 )}{\Delta_{ik}^2} \Delta_{il}\Delta_{ij} = \sum_k \frac{\sin^2 (\Delta_{i k} t /2 )}{\Delta_{ik}^2} \left|\sum_j V_{ij}V_{jk} \Delta_{ij}\right|^2
\end{equation}
When $\Delta_{ik}$ is close to zero, $\Delta_{ij} \approx -\frac{1}{2} (\Delta_{jk} - \Delta_{ij})$, and so $\sum_{j} V_{ij}V_{jk}\Delta_{ij} \approx -\frac{1}{2}[V,[H_0,V]]_{ik}$. Treating the sinc as proportional to a delta function as before, we have a contribution of 
\begin{equation}
    \frac{d}{dt} G^{(4)}(t)\big|_{\text{FGR}} \sim \frac{\pi^5 \lambda^4 }{32 \omega^4} \mathcal{A}_i^{[V,[H_0,V]]}(0)
\end{equation}

We also have a contribution from $G^{(3)}$ which stems from cross-terms between the $[H_0,V]$ and $[V, [H_0,V]]$ terms. We have, taking $V$ real for simplicity, and keeping only terms of the relevant order in $\frac{\lambda}{\omega}$,
\begin{equation}
\begin{split}
    \Re\left( \langle i |U^{(3)}_I |i \rangle\right) &= \frac{\pi^4\lambda^4}{12\omega^4}  \sum_{jk}\left( \frac{\sin^2 (\Delta_{i j} t /2 )}{\Delta_{ij}^2 \Delta_{kj}}+ \frac{\sin^2 (\Delta_{i k} t /2 )}{\Delta_{ik}^2 \Delta_{jk}}\right) \Bigg(\Delta_{ij}\Delta_{jk} V_{ij}V_{jk}[V,[H_0,V]]_{ki} + \\&
    \Delta_{jk}\Delta_{ki} [V,[H_0,V]]_{ij}V_{jk}V_{ki} + \Delta_{ij}\Delta_{ki} V_{ij}[V,[H_0,V]]_{jk}V_{ki} \Bigg)  
\end{split}
\end{equation}
Through relabeling of $j\leftrightarrow k$, this becomes
\begin{equation}
\begin{split}
    \Re\left( \langle i |U^{(3)}_I |i \rangle\right) = \frac{\pi^4\lambda^4}{6\omega^4}  \sum_{jk}\frac{\sin^2 (\Delta_{i k} t /2 )}{\Delta_{ik}^2 \Delta_{jk}} \Bigg(\Delta_{ik}\Delta_{kj} [V,[H_0,V]]_{ij}V_{jk}V_{ki} + 
    \Delta_{kj}\Delta_{ji} V_{ij}V_{jk}[V,[H_0,V]]_{ki} + \Delta_{ik}\Delta_{ji} V_{ij}[V,[H_0,V]]_{jk}V_{ki} \Bigg)  
\end{split}
\end{equation}
Note that the late-time contribution of the sinc term comes from $\Delta_{ik}$ close to $0$,  which eliminates the first and last terms in the parentheses under Fermi's Golden Rule. That is, our late-time behavior is determined by the middle term,
\begin{equation}
\begin{split}
    \Re\left( \langle i |U^{(3)}_I |i \rangle\right)|_\text{FGR} &= \frac{\pi^4\lambda^4}{6\omega^4}  \sum_{jk}\frac{\sin^2 (\Delta_{i k} t /2 )}{\Delta_{ik}^2 \Delta_{jk}} \Bigg( 
    \Delta_{kj}\Delta_{ji} V_{ij}V_{jk}[V,[H_0,V]]_{ki}\Bigg)\\
    &= \frac{\pi^4\lambda^4}{6\omega^4}  \sum_{k} [V,[H_0,V]]_{ik} \frac{\sin^2 (\Delta_{i k} t /2 )}{\Delta_{ik}^2} 
    \sum_j \Delta_{ij} V_{ij}V_{jk}
\end{split}
\end{equation}
When $\Delta_{ik}$ is close to zero, $\Delta_{ij} \approx -\frac{1}{2} (\Delta_{jk} - \Delta_{ij})$, and so $\sum_{j} V_{ij}V_{jk}\Delta_{ij} \approx -\frac{1}{2}[V,[H_0,V]]_{ik}$, so we have, under the FGR assumptions,
\begin{equation}
\begin{split}
    \Re\left( \langle i |U^{(3)}_I |i \rangle\right)|_\text{FGR} \approx -\frac{\pi^4\lambda^4}{12\omega^4}  \sum_{k} |[V,[H_0,V]]_{ik}|^2 \frac{\sin^2 (\Delta_{i k} t /2 )}{\Delta_{ik}^2} 
\end{split}
\end{equation}
and
\begin{equation}
\begin{split}
     \frac{d}{dt} G^{(3)}(t)\big|_{\text{FGR}} = -\frac{\pi^5\lambda^4}{12\omega^4} \mathcal{A}_i^{[V,[H_0,V]]}
\end{split}
\end{equation}
Putting together the three contributions, we have a rate
\begin{equation}
    R^{\text{intra}}_i = \frac{\pi^5 \lambda^4}{\omega^4} \mathcal{A}_i^{[V,[H_0,V]]}(0) \left(\frac{1}{18}+\frac{1}{32}-\frac{1}{12} \right) = \frac{\pi^5 \lambda^4}{288 \omega^4} \mathcal{A}_i^{[V,[H_0,V]]}(0)
\end{equation}
This rate for the odd drive matches that of the even drive. 
\subsection{Crossover frequency}
In this section, we approximate the crossover frequency where the intrazone rate begins to dominate the interzone rate. We assume that the crossover $\omega_c$ occurs at a value of $\omega$ that is either comparable to or larger than the local energy scale $J$ of $H_0$; as we will see below, this occurs when $\lambda$ is small relative to $J$. For simplicity, we consider the sinusoidal drive with $g(t) = \sin(t)$, but the crossover frequency for a general drive will have the same parametric dependence on the energy scales $\lambda$ and $J$. The intrazone and interzone rates for this drive are
\begin{equation}
    R^{\text{intra}}_i = \frac{\pi \lambda^4}{8 \omega^4} \mathcal{A}^{[V,[H_0,V]]}_i(0)
\end{equation}
\begin{equation}
    R^{\text{inter}}_i = \frac{\pi \lambda^2}{2}  \left(\mathcal{A}^V_i(\omega) + \mathcal{A}^V_i(-\omega) \right) \sim \pi \lambda^2 a^V_i e^{-\omega/J}
\end{equation}
In the above asymptotic expression for the interzone rate, we used the fact noted in the main text that $\mathcal{A}^V_i(\omega)$ decays exponentially with frequency \cite{Abanin-Huveneers2015_exponentially}. We wrote the asymptotic proportionality constant between $\mathcal{A}^V_i(\omega)$ and $e^{-\omega/J}$ as $a^V_i$, and we expect this constant to be approximated by or strongly correlated with $\mathcal{A}^V_i(0)$. Setting the asymptotic expressions equal to one another yields
\begin{equation}
 \left( \frac{\omega}{J} \right)^4 e^{-\omega/J} = \frac{\lambda^2 \mathcal{A}_i^{[V,[H_0,V]]}(0)}{8 J^4 a^V_i}
\end{equation}
The crossover frequency is the largest $\omega$ solution to this, above which the intrazone rate dominates. This is equivalent to solving the problem
\begin{equation}
    x^4 e^{-x} = k
\end{equation}
for the largest solution. Solving via iteration yields
\begin{equation}
\begin{split}
&x \sim \log(\frac{1}{k}) + 4\log(\log(\frac{1}{k})) \\
&e^{-x} \sim \frac{k}{\log(\frac{1}{k})^4}
 \end{split}
\end{equation}
so we have that the crossover frequency $\omega_i^c$ is 
\begin{equation}
    \omega_i^c \sim J \log \left( \frac{8 J^4 a^V_i}{\lambda^2 \mathcal{A}_i^{[V,[H_0,V]]}(0)} \right)
\end{equation}
and the rate $R_i^c$ at the crossover is 
\begin{equation}
    R_i^c \sim \frac{\pi \lambda^4 \mathcal{A}_i^{[V,[H_0,V]]}(0)}{8 J^4 \log \left( \frac{8 J^4 a^V_i}{\lambda^2 \mathcal{A}_i^{[V,[H_0,V]]}(0)} \right)^4}
\end{equation}
Note that $H_0$ implicitly contains a factor of $J$, and the averaging in the spectral function implicitly yields an inverse factor of $J$, so we have that $a_i^V \sim L^d/J$ and $ \mathcal{A}_i^{[V,[H_0,V]]}(0) \sim L^d J$. Neglecting $O(1)$ factors, the dominant dependence on $J$ and $\lambda$ are through $\omega_i^c \sim J \log(J^2/\lambda^2) \sim J \log(J/\lambda)$ and $R_i^c \sim L^d J \frac{(\lambda/J)^4}{\log(J/\lambda)^4}$ at small $\lambda$. Note that there are $O(1)$ factors that depend on the details of the state, $V$, and $H_0$ through the relative magnitude of off-diagonal matrix elements of $V$ and $[V,[H_0,V]]$. 

In one dimension, spectral functions of local operators decay weakly superexponentially in the frequency $\omega$ \cite{Abanin-Huveneers2015_exponentially}; this only changes the results above by multiplicative factors of $\log(\log(J/\lambda))$.

\section{Connections to U(1) prethermalization}
We noted in the main text that there is a close relationship between prethermalization at rapid drive frequencies $\omega$ and $U(1)$ prethermalization in the presence of a large magnetic field $h S^z$; this allows us to generalize our results on eigenstate stability.

$U(1)$ prethermalization involves a Hamiltonian $H=H_0+hS^z+\lambda V$ with $H_0$ conserving $U(1)$ and $V$ breaking $U(1)$. Despite the fact that the full Hamiltonian $H$ no longer conserves $U(1)$, the $U(1)$ charge looks conserved on times that grow exponentially in $h$. This mimics the exponentially long heating timescales of a rapidly driven Hamiltonian, and there is a mapping between the two problems discussed in ~\cite{Abanin-Huveneers2017_rigorous}.

We can use an analogy between Floquet zones and magnetization sectors to extend our results on the stability of eigenstates to time-independent systems demonstrating $U(1)$ prethermalization. A perturbation like $V=S^x$ induces ``interzone" and ``intrazone" processes between and within magnetization sectors of $H_0$ in time-dependent perturbation theory. 

Note that matrix elements $V_{ij}$ that couple eigenstates of $H_0 + h S^z$ in magnetization sectors $m$ and $m+\delta$ are coupling eigenstates of $H_0$ at energies separated by $h \delta$; these matrix elements are correspondingly exponentially suppressed in $h$ ~\cite{Abanin-Huveneers2015_exponentially}. That is, magnetic interzone processes are exponentially suppressed in $h$ in a close parallel to the driven interzone processes.

However, there is generically a difference that affects the strength of the intrazone channel: in the driven case, the matrix elements coupling a Floquet zone to its neighbor at a harmonic $p\omega$ above and to its neighbor at a harmonic $-p\omega$ below are guaranteed to be the same. This meant that terms contributing to driven intrazone processes that were naively of order $O(\frac{1}{\omega^2})$ largely cancelled and were suppressed down to order $O(\frac{1}{\omega^4})$ (see Eq.~8 in the main text; see also Eq.~\ref{Eq:cancellation} and surrounding discussion).

In the $U(1)$ prethermalization case without additional conditions, matrix elements $V_{ij}$ coupling eigenstates of $H_0$ within $S^z=m$ to those within $S^z=m+\delta$ need not be identical to matrix elements coupling $S^z=m$ to $S^z=m-\delta$. That is, more conditions are needed to guarantee the cancellation of processes contributing to the magnetic intrazone channel.

For example, this cancellation is guaranteed for an initial state in the $S^z=0$ sector of $H_0$ in the presence of a spin-flip $S^z \leftrightarrow -S^z$ symmetry generated by $U=e^{i \pi S^x}$; specifically, it requires $U^\dagger H_0 U = H_0$ and $U^\dagger V U = \pm V$. When these conditions are met, the magnetic intrazone processes undergo a similar cancellation to the driven intrazone processes with an asymptotic timescale scaling as $h^4$. 

More generically, when these conditions are not met, the magnetic intrazone processes lead to a faster decay timescale scaling as $h^2$. However, in either case, these asymptotic regimes can first be preceded by the interzone timescale growing exponentially in $h$.

\bibliography{main}